\pdfoutput=1
\documentclass[aps,prd,12pt,nofootinbib]{revtex4}
\usepackage{epsfig}
\usepackage{graphicx}
\usepackage[font=small,skip=0pt,justification=raggedright]{caption}
\usepackage{amsmath}
\usepackage{amssymb}
\usepackage{amsfonts,amsthm}
\usepackage{mathrsfs}
\usepackage{verbatim}
\usepackage{dutchcal}

\usepackage[normalem]{ulem}
\usepackage{xcolor}

\newcounter{fig}   \newcommand{\lbfig}[1]{\refstepcounter{fig}
\label{#1} }
\newcommand{\vphi}{\varphi}

\newcommand{\bea}{\begin{eqnarray}}
\newcommand{\eea}{\end{eqnarray}}
\newcommand{\be}{\begin{equation}}
\newcommand{\ee}{\end{equation}}

\newcommand{\re}[1]{(\ref{#1})}

\newcommand{\ta}{\theta}

\newcommand{\pa}{\partial}

\newcommand{\eqn}{\begin{eqnarray}}
\newcommand{\eqnx}{\end{eqnarray}}

\begin{document}
\title{$U(1)$ gauged non-topological solitons
in the 3+1 dimensional $O(3)$ sigma-model}
\author{L. A. Ferreira}
\affiliation{Instituto de F\'{i}sica de S\~{a}o Carlos; IFSC/USP;
Universidade de  S\~{a}o Paulo, USP\\
Caixa Postal 369, CEP 13560-970, S\~{a}o Carlos-SP, Brazil}
\author{A. Mikhaliuk}
\affiliation{Belarusian State University, Minsk 220004, Belarus}
\author{Y. Shnir}%
\affiliation{BLTP, JINR, Dubna 141980, Moscow Region, Russia}

\begin{abstract}

We present and study new non-topological soliton solutions in the
$U(1)$ gauged  non-linear $O(3)$ sigma model with a symmetry breaking
potential in 3+1 dimensional flat space-time. The configurations are endowed with an
electric and magnetic field and also carry a nonvanishing angular momentum density. We discuss properties of these solitons and
investigate the domains of their existence.

\end{abstract}
\maketitle
 \section{Introduction } \label{Introduction}


The nonlinear $O(3)$ sigma-model has been proposed long time ago by
Gell-Mann and Levy \cite{Gell-Mann:1960mvl} in the context of a theoretical description of the
low-energy dynamics of  pions.
Nowadays, it is considered as a well-known prototype for a large class of  field theories
supporting topological solitons, see e.g. \cite{Manton:2004tk,Shnir:2018yzp}. The field of the
$O(3)$ sigma-model is restricted to a sphere $S^2$, in 2+1 dimensions it is a topological map $S^2 \mapsto S^2$ which belongs to an equivalence class characterized by the homotopy
group $\pi_2(S^2)=\mathbb{Z}$. Hence, there are
soliton solutions of the model \cite{Polyakov:1975yp}.
Further, in 3+1 dimensions, the field configurations of the nonlinear $O(3)$ sigma-model define a map
$\mathbb{R}^3 \mapsto S^2$,  which is characterised by the third homotopy group $\pi_3(S^2)$.

However, in 3+1 dimensions, the Derrick's theorem \cite{Derrick:1964ww} does not allow for
the existence  of such static finite energy solitons in the usual $O(3)$
sigma model, which includes only the quadratic in derivative terms. Such topological solitons, referred as Hopfions,
appear in the scale-invariant Nicole model \cite{Nicole:1978nu} and, more importantly, in the
Faddeev-Skyrme model \cite{Faddeev:1976pg}, which
contains terms with both two and four derivatives\footnote{
There is a vast literature on solitons in 2+1 dimensional nonlinear $O(3)$ sigma-model and its generalizations,
which we will not discuss here.}. Apart from the addition of higher derivative terms, no
other mechanism to secure stability of the regular topological solitons of the $O(3)$
sigma model in three spacial dimensions is known. Inclusion of the potential term \cite{Foster:2010zb} alone
cannot stabilize the configuration, as well as gauging of the  $U(1)$ subgroup of the
$O(3)$ symmetry and including the Maxwell term in the total Lagrangian \cite{Shnir:2014mfa,Samoilenka:2018oil}.

The global symmetry of the $O(3)$ sigma model admits internal rotations of the
components of the real scalar field, it may allow to circumvent the restrictions of Derrick's
theorem \cite{Ward:2003un}. So-called "Q-lumps" in 2+1 dimensions are isorotating solitons which
carry both Noether and topological charges \cite{Leese:1991hr,Ward:2003un}

However, isorotations of the Hopfions are allowed only in the presence
of both potential and higher derivative terms \cite{Harland:2013uk,Battye:2013xf}. On the other hand,
a stable isospinning non-topological solitons, Q-balls, can exist in
a general class of scalar field theories with a global $U(1)$ symmetry \cite{Rosen:1968mfz,Friedberg:1976me,Coleman:1985ki},
in particular, in the non-linear $O(3)$ sigma model with a non-negative potential \cite{Verbin:2007fa}.
Nonetheless, it was pointed out that the usual weakly attractive "pion mass" potential \re{U-pion} cannot stabilize isorotating $O(3)$
configurations in 3+1 dimensions \cite{Herdeiro:2018djx}, unless gravitational attraction is included.
However, the virial theorem does not exclude existence of regular isospinning $O(3)$ solitons stabilized by internal rotation
in the flat space \cite{Herdeiro:2018djx},
certain modification of the potential term may allow for construction of such non-topological
solitons in 3+1 dimensional space \cite{Verbin:2007fa,Adam:2025ktm}.

Further, the global $U(1)$ symmetry of a model supporting Q-balls
can be promoted to a local gauge symmetry, with the corresponding gauged Q-balls possessing  electric charge
\cite{Rosen:1968mfz,Lee:1988ag,Anagnostopoulos:2001dh,Gulamov:2013cra,Gulamov:2015fya}.
The presence of the long-range gauge field affects the properties of the solitons, as the gauge
coupling increases, the electromagnetic repulsion may destroy configurations.

The objective of this paper is to analyze properties of the regular solutions of the
$U(1)$ gauged  non-linear $O(3)$ sigma model with modified potential in 3+1 dimensional flat space,
focusing our study on non-topological localized configurations
of different types, and determine their domains of existence.

This paper is organized as follows. In Section II, we introduce the model,
and the field equations with the stress-energy tensor of the system of interacting fields.
Here we describe the axially-symmetric parametrization of the matter fields and
the boundary conditions imposed on the configuration.
We also discuss the physical quantities of interest. In Sec. III we present the
results of our study with particular emphasis on the role of the electromagnetic interaction.
We conclude with a discussion of the results and final remarks.

\section{The model}
\subsection{The action and the field equations}
We consider the $U(1)$ gauged non-linear sigma-model in a
(3+1)-dimensional Minkowski space-time. The corresponding matter field Lagrangian of the system is
\be
L_{m} = -\frac14 F_{\mu\nu}F^{\mu\nu} + \frac12 (D_\mu\phi^a)^2 - U(\phi) \, ,
\label{Gauged_O3}
\ee
where the real triplet of the scalar fields $\phi^a$, $a=1,2,3$, is restricted to
the surface of the
unit sphere, $(\phi^a)^2=1$, and $U(\phi)$ is a symmetry breaking potential. In particular, we can consider
simple "pion mass" potential
\be
U(\phi)=\mu^2 (1 -\phi^3),
\label{U-pion}
\ee
where a mass parameter $\mu$ is a positive constant.

The $U(1)$ field strength tensor is
$F_{\mu\nu}=\partial_\mu A_\nu-\partial_\nu A_\mu$,
and the covariant derivative of the  field $\phi^a$, that minimally
couples the scalar field to the gauge potential is
\be
D_\mu \left(\phi^1+i\,\phi^2\right) = \partial_\mu \left(\phi^1+i\,\phi^2\right)
-i\,e\,A_\mu\left(\phi^1+i\,\phi^2\right)\,, \qquad D_\mu
\phi^3 = \partial_\mu \phi^3,
\label{covariant}
\ee
with $e$ being the gauge coupling. The vacuum boundary conditions are
$\phi_{\infty}^a = (0,0,1),\, D_\mu\phi^a =0,\, F_{\mu\nu}=0$. In the static
gauge, where the fields have no explicit dependence on time, the
asymptotic boundary conditions on the gauge potential are
\be
A_0(\infty)=V,\quad A_i(\infty)=0
\ee
where $V$ is a real constant.
The electromagnetic and scalar components of the energy-momentum tensor are, respectively
\be
\begin{split}
T_{\mu\nu}^{Em} &=F_\mu^\rho F_{\nu\rho} - \frac14 \eta_{\mu\nu} F_{\rho\sigma}  F^{\rho\sigma}\,,\\
T_{\mu\nu}^{\phi} &= D_\mu\phi^a D_\nu\phi^a
 - \eta_{\mu\nu}\left[\frac{\eta^{\rho\sigma}}{2}  D_\rho\phi^a
D_\sigma\phi^a +  U(\phi)\right] \, ,
\label{Teng}
\end{split}
\ee
where $\eta_{\mu\nu}=~{\rm diag}\, (-1,1,1,1)$ is the Minkowski metric.

The potential $U(|\phi|)$ breaks $O(3)$ symmetry to the subgroup $SO(2)\sim U(1)$,
the model \re{Gauged_O3} is invariant with respect to
the local Abelian gauge transformations
\be
(\phi^1 + i \phi^2) \to
e^{ie \zeta}(\phi^1 + i \phi^2), \quad A_\mu \to A_\mu + \partial_\mu \zeta
\label{rotate}
\ee
where $\zeta$ is a real function of the coordinates. The corresponding Noether current
can be written as follows:
\be
j^\mu=\phi^1 D^\mu\phi^2 - \phi^2 D^\mu\phi^1
\label{ncurrent}
\ee

Variation of the Lagrangian \re{Gauged_O3}  with respect to
the fields $A_\mu$ and $\phi_a$ leads to the equations:
\be
\partial_\mu  F^{\mu\nu}=e  j^\nu,
\quad D^\mu D_\mu \phi^a + \phi^a( D^\mu \phi^b  D_\mu \phi^b)
+ \mu^2\left[ \phi^a(\phi^b \phi^b_\infty) -\phi^a_\infty\right]=0\, ,
\label{eqfield}
\ee
where we take into account dynamical constraint imposed on the $O(3)$ field and make use of the explicit form of the potential
\re{U-pion}.

We remark that asymptotic value
of the electric potential $A_0(\infty)$
can be adjusted via the residual $U(1)$ transformations, choosing  $\zeta =-Vt$.
In the stationary gauge $A_0(\infty)=0$ and two components \re{rotate} of the scalar
triplet acquire an explicit harmonic time  dependence with frequency $\omega=eV$.

\subsection{Virial identity}
The Derrick's theorem \cite{Derrick:1964ww} indicates that the usual non-linear $O(3)$ sigma-model
do not possess stable solitonic solutions in  $3+1$ dimensional flat space. By contrast, in the
$U(1)$ gauged model \re{Gauged_O3},  the usual scaling arguments  suggest that localized regular solutions
may be allowed.  The total energy functional of the system can be written as
\be
E= E_2 + E_0 + E_4 \, ,
\label{eng}
\ee
where
\be
E_2=\frac12\int d^3x  D_i\phi^a D_i\phi^a ,\quad
E_0= \int d^3x  U(\phi), \quad
E_4=\frac12 \int d^3x (E_i^2 + B_i^2)\, ,
\label{Derek}
\ee
and $E_i=F_{0i}$ and $B_i=\frac12 \varepsilon_{ijk}F_{jk}$ are the electric and magnetic fields, respectively.

The critical points of the total energy functional \re{eng} should satisfy the arguments of Derrick's theorem.
The scale transformation $x^i\rightarrow {x^i}^\prime = \lambda^{-1} x^i$, do not affect the scalar field,
$\phi^a \rightarrow \phi^a$ because of the sigma model constraint,
and $A_\mu(x_i) \rightarrow A_\mu^\prime(x_i^\prime) =\lambda\, A_\mu(x_i)$. Then
\be
E(\lambda) =\lambda^{-1}E_2 + \lambda E_4 +
\lambda^{-3}E_0
\ee
and $\partial^2_\lambda E(\lambda=1)=2(E_2+6E_0) \ge 0$. The corresponding virial identity
follows from the condition $\partial_{\lambda} E\mid_{(\lambda=1)}=0$, it gives
\be
E_2+3E_0=E_4 \, ,
\label{virial}
\ee
which suggests a possibility of existence of soliton solutions stabilized by electromagnetic field
\cite{Schroers:1995he},
see also \cite{Amari:2024adu,Loginov:2016yeh,Ghosh:1995ze,Samoilenka:2015bsf}.

We also note that non-topological solitons of the model \re{Gauged_O3} in Minkowski
space may exist also in the limiting case of vanishing electromagnetic interactions since the
internal rotations with a non-zero angular frequency $\omega$ stabilize the configuration in the
same way as it happens for the usual scalar Q-balls \cite{Ward:2003un}.
Indeed, the total energy of a stationary spinning configuration also contains the term $E_{\omega}=\omega^2 \Lambda/2$, where $\Lambda=\int d^2 x ((\phi^1)^2+(\phi^2)^2 $ is the moment of inertia. This term yields an effective (tachionic) mass to the
field $f$.  Then, under the scaling transformations in $d=3$
\be
E(\lambda) =\lambda^{-1}E_2 + \lambda E_4 +
\lambda^{-3}E_0 + \lambda^{3}E_\omega
\ee
and
the corresponding virial relation becomes
$$
E_2+3E_0 = \frac{3}{2}\omega Q \, .
$$
where $Q$ is the Noether charge corresponding to the current \re{ncurrent}.

However, neither isorotations \cite{Herdeiro:2018djx} nor electromagnetic interactions can support
stable non-topological solitons in $O(3)$ sigma model.  Eventually, there exists yet another loophole in the no-go arguments based on the Derrick's theorem. The  scalar potential  may be taken  as the difference of positive definite terms, $U(\phi) = U(\phi)^{(1)}-U(\phi)^{(2)}$, then, in the absence of the Maxwell term and isorotations,  the  corresponding virial identity becomes
\be
E_2+3E_0^{(1)}=3E_0^{(2)} \, ,
\ee
where both $E_0^{(1)}$ and $E_0^{(2)}$
are positive-definite. In order to secure existence of solitonic solutions of the model
\re{Gauged_O3}  the underlying potential also must possess two minima  of different depths
\cite{Verbin:2007fa}. Both electromagnetic interactions and isorotations may further affect such solution.

\subsection{$\mathbb{C}P^1$ formulation}

In order to gain some insight into the construction of soliton solutions of the model \re{Gauged_O3} in the flat space-time, we
make use of inhomogeneous
coordinates on $\mathbb{C}P^1$,
\be
u=\frac{\phi^1 + i\phi^2}{1-\phi^3};\qquad \qquad {\vec \phi}=\frac{1}{1+|u|^2}\left(u+u^*\,,\,-i(u-u^*)\,,\,|u|^2-1\right) \, .
\ee
Then  we can recast\footnote{Note that $L=L_m/2$.} the Lagrangian \re{Gauged_O3} in  terms of the $\mathbb{C}P^1$ fields
\be
L= -\frac18 F_{\mu\nu}F^{\mu\nu} + \frac{(D_\mu u) (D^\mu u)^*}{(1+|u|^2)^2}  - U(|u|) \, .
\label{Gauged_O3-u}
\ee
Here $D_\mu u = \pa_\mu u -ie A_\mu u$,  under the $U(1)$ gauge transformation \re{rotate}
$u\to e^{ie \zeta(x)}u$,  $A_\mu \to A_\mu + \partial_\mu \zeta  $, the derivative $D_\mu u$ transforms covariantly.

Variation of the Lagrangian \re{Gauged_O3-u} with respect to the  field $u^*$ yields the equation
\be
(1+|u|^2)D_\mu D^\mu u - 2\,u^*\, (D_\mu u) (D^\mu u)  + u\,(1+|u|^2)^3\, \frac{\delta U}{\delta |u|^2} =0 \, .
\label{u-eqs}
\ee
The corresponding equation for the electromagnetic field is
\be
\partial_\mu F^{\mu\nu}=e\, j^\nu \, ,
\label{F-eqs}
\ee
where the conserved current \re{ncurrent} is rewritten as
\be
j^\nu= 2\,i \frac{\left[u (D^\nu u)^* - u^* (D^\nu u)\right]}{(1+|u|^2)^2} \, .
\ee

To construct solutions of the equations \re{u-eqs},\re{F-eqs}, we may employ the polar
decomposition of the complex field
\be
u= Fe^{i\Theta} \, ,
\ee
where both the phase  $\Theta \in [0,2\pi]$  and  the amplitude $F\in [0,\infty]$ are real
functions of the coordinates in
3+1 dimensional space-time. Note that the original triplet of
real scalar fields then can be written as
$$
\vec \phi =\left(\frac{2F }{1+F^2}\cos \Theta,\, \frac{2F }{1+F^2}\sin \Theta,\, \frac{F^2-1}{F^2+1} \right)
$$
Now, denoting $F=\cot \frac{f}{2}$, $f\in [0,\pi]$, we arrive at the following "trigonometric"
parametrization of the $O(3)$ field
\be
{\vec \phi} =\left(\sin f \cos \Theta,\,
\sin f \sin \Theta,\, \cos f
\right) \, .
\label{trig-phi}
\ee

In terms of the trigonometric parametrization \re{trig-phi}, the real and imaginary parts of the equation of motion \re{u-eqs} leads to the following two equations
\be
\partial_{\mu}^2 f - e^2 \sin f\,\cos f\, C_{\mu}^2+\frac{\delta U}{\delta f} \, .
\label{eqforf}
\ee
and
\be
\partial^{\mu}\left(\sin^2 f\, C_{\mu}\right)=0 \, .
\label{eqforfcmu}
\ee
with $C_{\mu}$ being the gauge invariant vector field
\be
C_{\mu}\equiv A_{\mu}-\frac{1}{e}\,\partial_{\mu} \Theta \, .
\ee
The equation for the electromagnetic field \re{F-eqs} becomes
\be
\partial_{\mu}^2C^{\nu}-\partial^{\nu}\partial^{\mu}C_{\mu}+e^2\,\sin^2 f\, C^{\nu}=0 \, .
\label{eqforcmu}
\ee
Note that the equations \re{eqforf}, \re{eqforfcmu} and \re{eqforcmu} are the same equations that one would get from  the Lagrangian \re{Gauged_O3}, by taking into account the constraint ${\vec \phi}^2=1$.

We point out that a term proportional to $(\partial_{\mu}f)^2$ in \re{eqforf}, that would come from the non-linear term in \re{eqfield}, or equivalently from the second term in \re{u-eqs}, vanishes due to the structure of the trigonometric parametrization \re{trig-phi}.

\section{Results}
\subsection{Choice of the potential}
In the unitary gauge the phase function $\Theta$ can be gauged away by applying the transformation
\re{rotate}. However, we find it more convenient in the numerical analysis to keep the phase function.
Further, let us consider, as a special case, a simple spherically symmetric
ansatz parameterized in terms of the radial function $f(r)$ and
the harmonic phase function which depends on time as $\Theta=-\omega t$.
The only non-zero component of the gauge potential is a radially depending function $A_t(r)$, i.e.
\be
f\equiv f(r);\qquad e\,C_0\equiv e\,A_t(r)+\omega; \qquad C_i=0; \qquad i=1,2,3
\label{radialansatz}
\ee

Note that the time-dependency of the $O(3)$ field  disappears at the level of the dynamical
equations and the energy-momentum tensor \re{Teng} does not depend upon time. Indeed, after inserting this
ansatz into the field equations \re{eqforf}, \re{eqforfcmu} and \re{eqforcmu}, or equivalently \re{eqfield},
we obtain
\be
\begin{split}
f^{\prime \prime} &+ \frac{2f^\prime}{r} + (\omega + eA_t)^2\sin f \cos f - \frac{\delta U}{\delta f}
=0,\\
A_t^{\prime \prime} &+
\frac{2A_t^\prime}{r}-e(\omega + eA_t)\sin^2 f =0\, ,
\end{split}
\label{sys-sph-flat}
\ee
where a prime denotes the radial derivative and the symmetry breaking potential $U(f)$
depends on the amplitude of the field. Note that our ansatz imply that $\partial_{\mu}C^{\mu}=0$ and $C^{\mu}\partial_{\mu}f=0$. Therefore, the equation \re{eqforfcmu} is automatically satisfied by the ansatz, and the middle term of \re{eqforcmu} vanishes.

 The simple
spherically-symmetric ansatz  \re{radialansatz} is self-consistent, as the same equations can be derived via
variation of the reduced Lagrangian
$$
L_{eff}=-\frac12 ( A_t^\prime)^2 -\frac12(\omega + eA_t)^2 \sin^2 f + \frac12 (f^\prime)^2 - U(f)\,
$$
with respect to the functions $f$ and $A_t$.

Choice of the explicit form of the symmetry breaking potential $U$ has a dramatic effect on the solutions of the model
\re{Gauged_O3}. Considering, for example, the usual weakly attractive "pion mass" potential \re{U-pion}, we
obtain two linearized asymptotic equations for the perturbative excitations of the fields around the vacuum
$f(r) \to 0 + \rho(r) + \dots,\quad A_t(r) \to  0 + a(r) + \dots
$,
\be
\rho^{\prime \prime} - \left(
\mu^2 -  \omega^2\right) \rho =0\, ,\qquad
a^{\prime \prime}  =0
\label{decay-eqs}
\ee
Hence, a localized configuration of the $O(3)$ field  with an exponentially decaying tail may
exist if $\omega \le \mu$. On the other hand, in the 3+1 dimensional flat space-time, the electric potential
remains massless.

\begin{figure}[t!]
\begin{center}
\includegraphics[height=.32\textheight,  angle =-90]{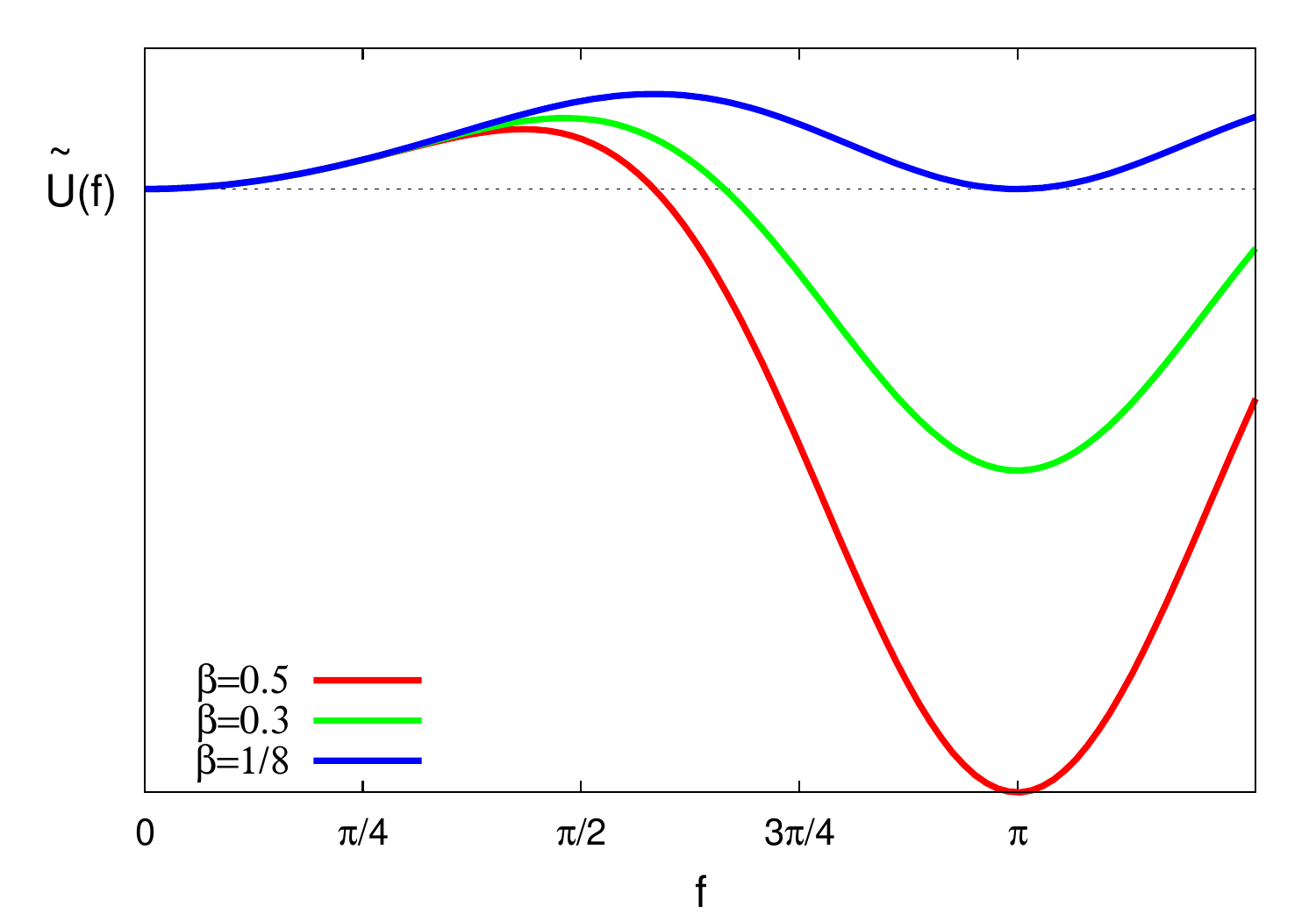}
\end{center}
\caption{\small The potential \re{pot-mod}
is shown for some set of values of the parameter $\beta$ and $\mu^2=1$.
}    \lbfig{fig1}
\end{figure}

However, numerical calculations indicate that for the  "pion mass" potential \re{U-pion} the virial identity
\re{virial} cannot be satisfied since the electrostatic energy always
remains smaller than the energy of the $O(3)$ scalar field, cf. related discussion in \cite{Herdeiro:2018djx}.

The situation changes dramatically when we consider a modified symmetry breaking potential, see Fig.~\ref{fig1}
\be
{\tilde U}(\phi)=\mu^2 (1 -\phi^3) - \beta (1-\phi^3)^4 \equiv \mu^2 (1 -\cos f) - \beta (1-\cos f)^4\,  ,
\label{pot-mod}
\ee
where $\beta $ is a real positive parameter. The potential \re{pot-mod}, apart from the usual "pion mass" term,
also contains a term, which yields an additional short-range interaction energy
\cite{Leese:1989gi,Salmi:2014hsa}. The corresponding  additional term in the field
equation decays faster than the contribution of the pion mass term, thus, the asymptotic equation
\re{decay-eqs}  remains the same. Note that similar choice of potential (with $\beta <0$) has also been used
in  \cite{Samoilenka:2015bsf,Salmi:2014hsa,Gillard:2015eia} to construct multisoliton solutions of the
Skyrme model with low binding energy.

Clearly, the modified potential $\tilde U(f)$ \re{pot-mod} displayed in  Fig.~\ref{fig1},
may support non-topological flat-space soliton solutions of the $U(1)$ gauged $O(3)$ sigma model.
It has a false minimum at $f=0$ and a global minimum at $f= \pi$. Here we stress an important
difference between the non-topological solitons in the theory of a complex scalar field
\cite{Coleman:1985ki,Lee:1991ax} and solutions of the
$O(3)$ model with the symmetry breaking potential
\re{pot-mod}: in the former case the true vacuum corresponds to the zero value of the field amplitude.

The depths of the global minimum at $f=\pi$ depends on the ratio of the  values of the parameters $\beta/\mu^2$,
and for $\beta/\mu^2= 1/8$ the vacuum becomes double degenerated.
The soliton solution then represent an infinitely thin spherically-symmetric wall
separating two neighboring vacua $\tilde U=0$ at $f_0^{(1)}=0$ and $f_0^{(2)}= \pi$.

\subsection{Ansatz and the boundary conditions}
Non-topological solitons of the gauged $O(3)$ model are akin to the usual Q-balls in
3+1 dimensional models of complex scalar field with a sextic potential
\cite{Coleman:1985ki,Lee:1991ax}. Indeed, using the spherically-symmetric ansatz
\re{trig-phi} with $\Theta=-\omega t$, we can find regular finite energy solutions
of the corresponding system of field equations \re{sys-sph-flat} displayed  in
Fig.~\ref{fig0}. In the time-dependent gauge both profile functions are asymptotically
vanishing, $f(\infty)=A_t(\infty)=0$, the Neumann boundary conditions at the origin are $\pa_r f(0)=\pa_r A_t(0) =0$.

Axially-symmetric solutions of the model \re{Gauged_O3} can be constructed using the stationary ansatz with a
harmonic time dependence, which is  similar  to the corresponding
parametrization of the  $O(4)$ Skyrme model
\cite{Battye:2014qva,Ioannidou:2006nn,Perapechka:2017bsb,Herdeiro:2018daq}
\be
{\vec \phi}=\left(\sin{f(r,\ta)}\cos{(n\vphi-\omega t)},\text{ }\sin{f(r,\ta)}\sin{(n\vphi-\omega t)},\text{ }\cos{f(r,\ta)}\right).
    \label{anssfield}
\ee
where the profile function $f$ depends on the  radial coordinate  $r$ and the polar angle $\ta$.
In the spherically symmetric case ($n=0$) this ansatz is reduced to \re{trig-phi}.
The corresponding ansatz for the $U(1)$ potential
contains two real functions of an electric and a magnetic potential:
\be
    A_\mu dx^\mu=A_t(r,\ta)dt + A_\vphi(r,\ta)\sin{\ta}d\vphi \, .
    \label{ansempot}
\ee
Note that the ansatz \re{anssfield}-\re{ansempot} satisfies automatically the equation  \re{eqforfcmu}.
After substituting this ansatz, we get the following field equations:
\be
\begin{split}
    &\left(\partial_r^2+\frac2r\partial_r+\frac{1}{r^2}\partial_{\theta}^2+\frac{\cos{\theta}}{r^2\sin{\theta}}\partial_{\theta}\right)f+\frac12\left(\left(eA_t+\omega\right)^2-\frac{1}{r^2}\left(eA_\varphi+\frac{n}{\sin{\theta}}\right)^2\right)\sin{2f}=\frac{\delta U}{\delta f};\\
    &\left(\partial_r^2+\frac2r\partial_r+\frac{1}{r^2}\partial_{\theta}^2+\frac{\cos{\theta}}{r^2\sin{\theta}}\partial_{\theta}-\frac{1}{r^2\sin^2{\theta}}-e^2\sin^2{f}\right)A_{\varphi}=\frac{ne}{\sin{\theta}}\sin^2{f};\\
    &\left(\partial_r^2+\frac2r\partial_r+\frac{1}{r^2}\partial_{\theta}^2+\frac{\cos{\theta}}{r^2\sin{\theta}}\partial_{\theta}-e^2\sin^2{f}\right)A_t=e\omega\sin^2{f};
\end{split}
\label{eqs-ax}
\ee

The boundary conditions are  $A_t=A_\vphi=0$ and $f=0$ at spacial infinity and
$\partial_r A_t=A_\vphi=\partial_r f=0$ at the origin (in the stationary gauge).
The regularity of the solutions at the symmetry axis requires that
($n\neq 0$)
\be
\pa_{\ta}f\vert_{\ta=0,\pi}= \pa_{\ta}A_t\vert_{\ta=0,\pi}=
    A_\vphi\vert_{\ta=0,\pi}= 0\,.
    \label{boundang1}
\ee
\begin{figure}[t!]
\begin{center}
\includegraphics[height=.32\textheight,  angle =-90]{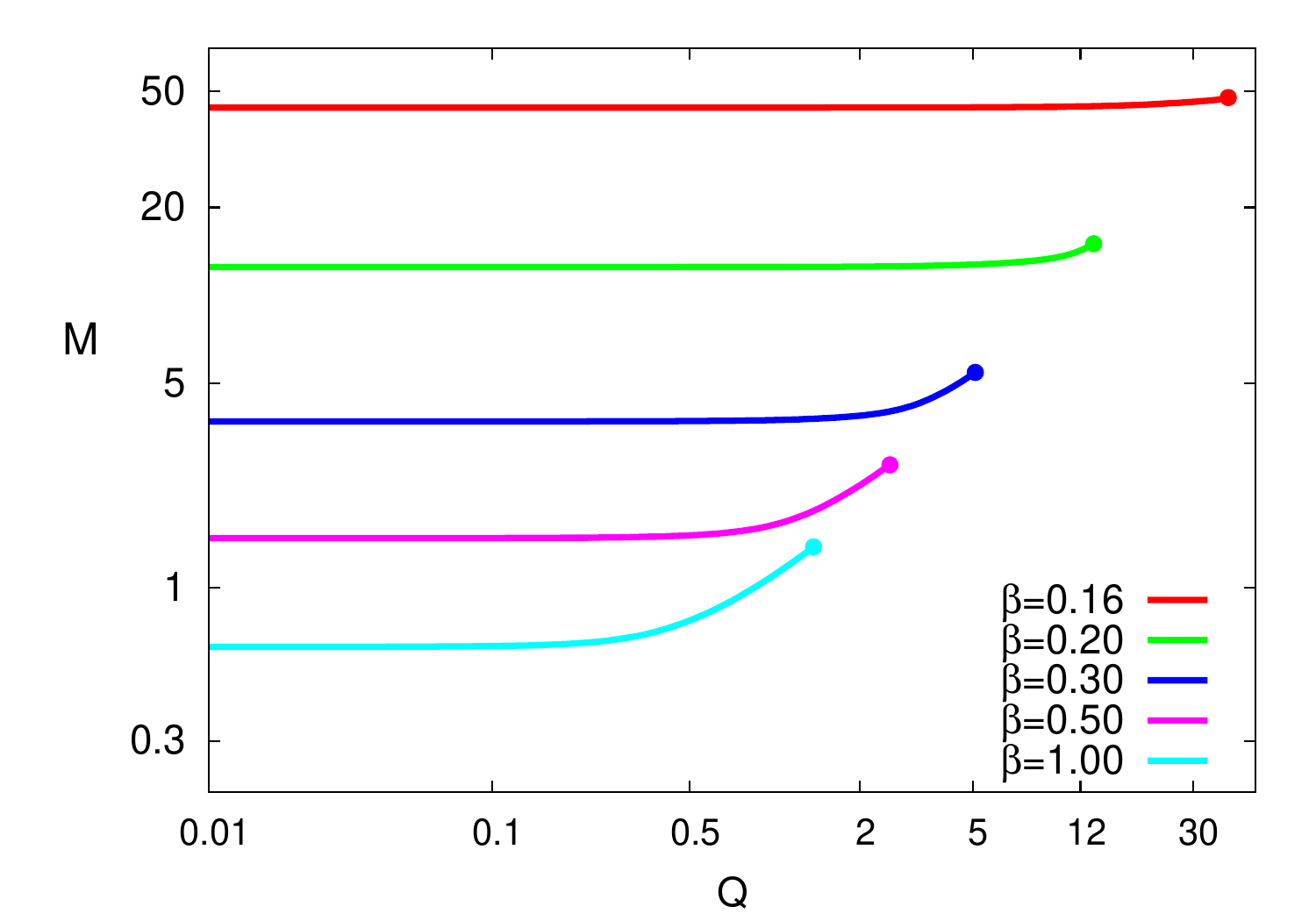}
\includegraphics[height=.32\textheight,  angle =-90]{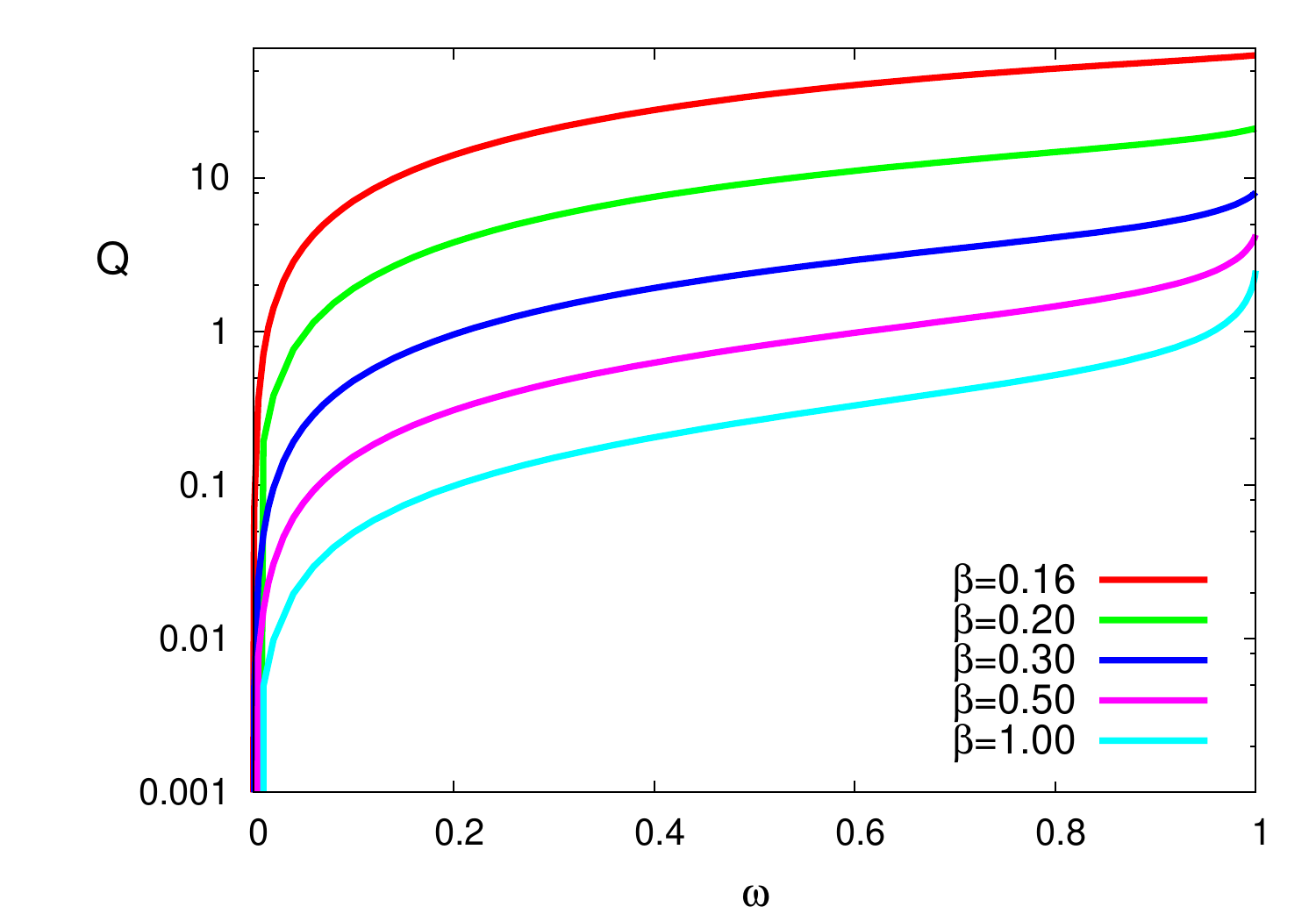}
\includegraphics[height=.32\textheight,  angle =-90]{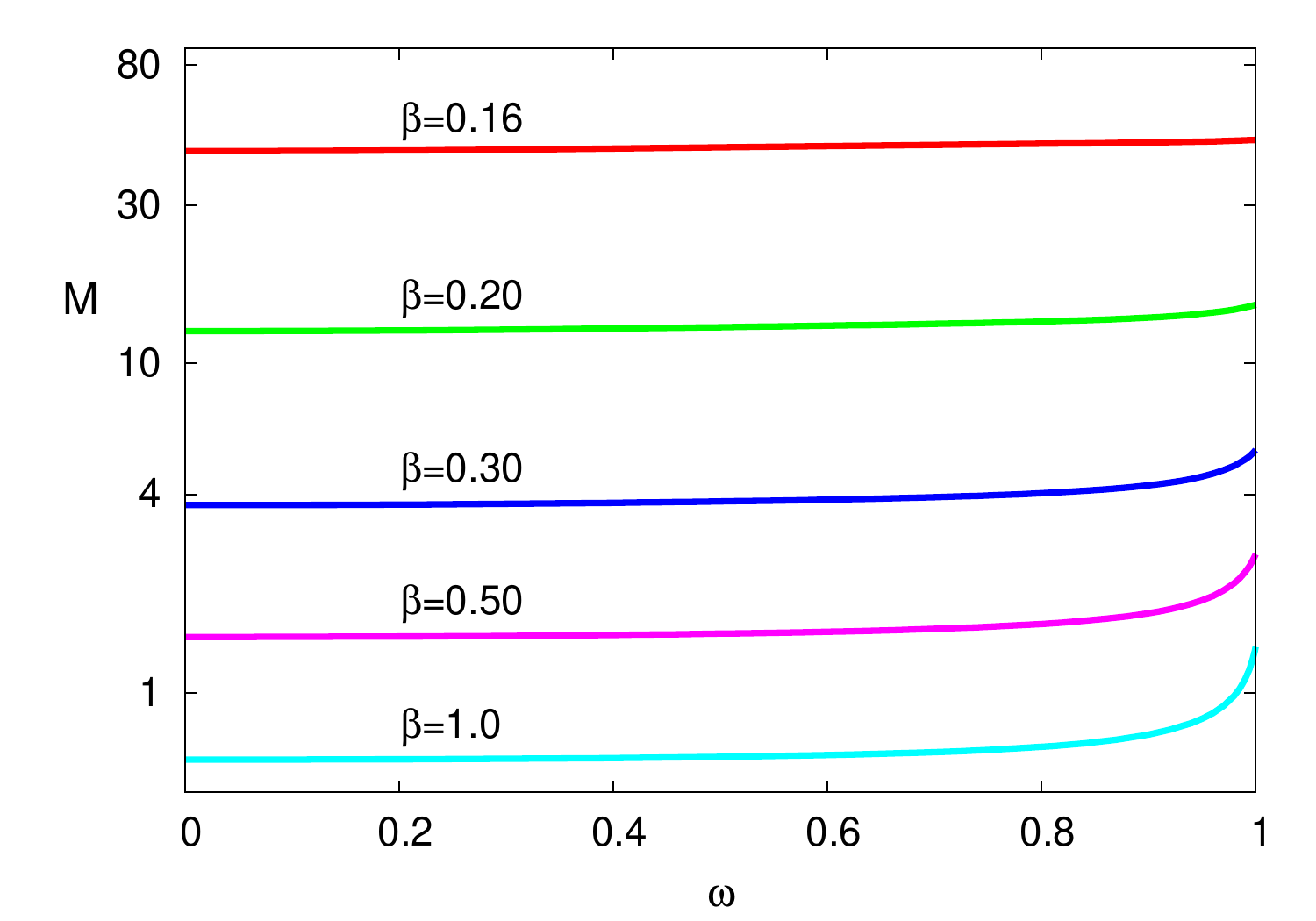}
\includegraphics[height=.32\textheight,  angle =-90]{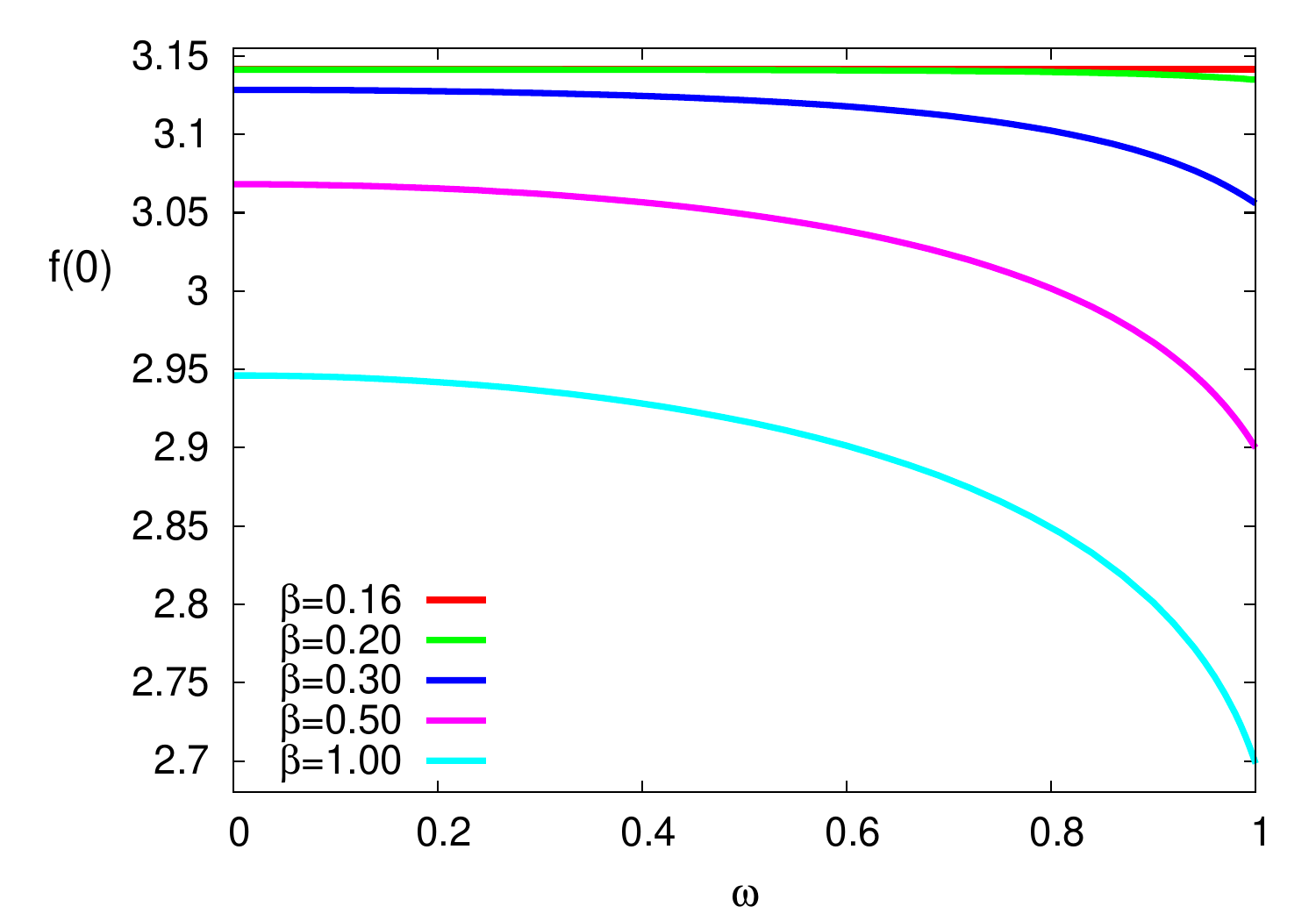}
\includegraphics[height=.32\textheight,  angle =-90]{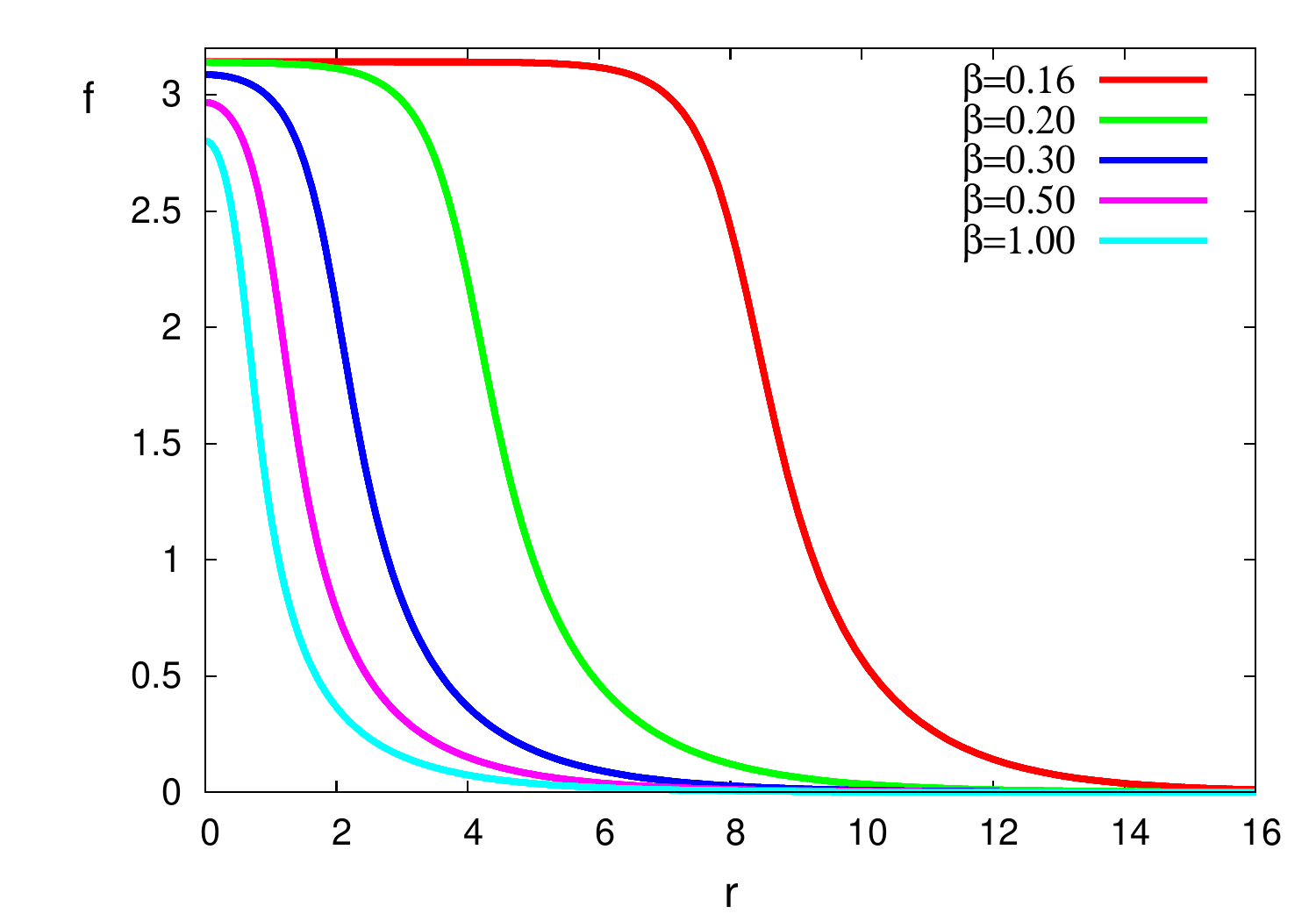}
\includegraphics[height=.32\textheight,  angle =-90]{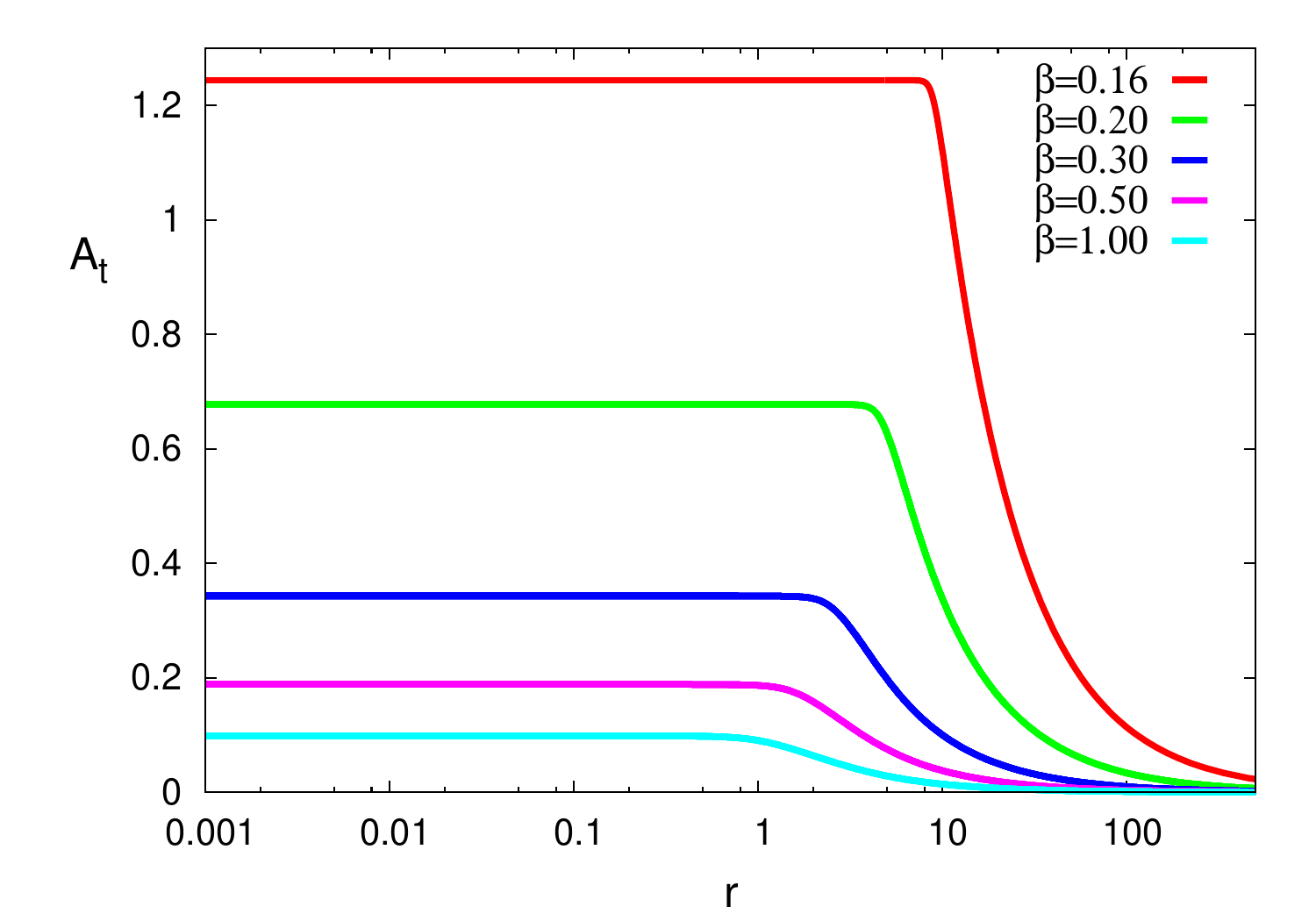}
\end{center}
\caption{\small Spherically symmetric $U(1)$ gauged $O(3)$ solitons:
Mass vs. charge $Q$ (upper left) and the charge $Q$ vs. frequency $\omega$
are shown for some set of values of the parameter $\beta$ and $\mu^2=1$. The dots on the upper left plot
indicate the limiting solutions with maximal
mass and charge at the threshold $\omega=\mu$. Middle plots: Mass $M$ (left) and the central value of the profile function $f(0)$ (right) vs the angular frequency $\omega$.
Bottom plots:
The radial profiles
of the fields $f(r)$ and $A_t(r)$  are displayed  for some set of values of the
parameter $\beta$ of the potential \re{pot-mod}  at frequency $\omega=0.90$, and gauge coupling $e=0.1$
}    \lbfig{fig0}
\end{figure}

\begin{figure}[t!]
\begin{center}
\includegraphics[height=.36\textheight,  angle =-90]{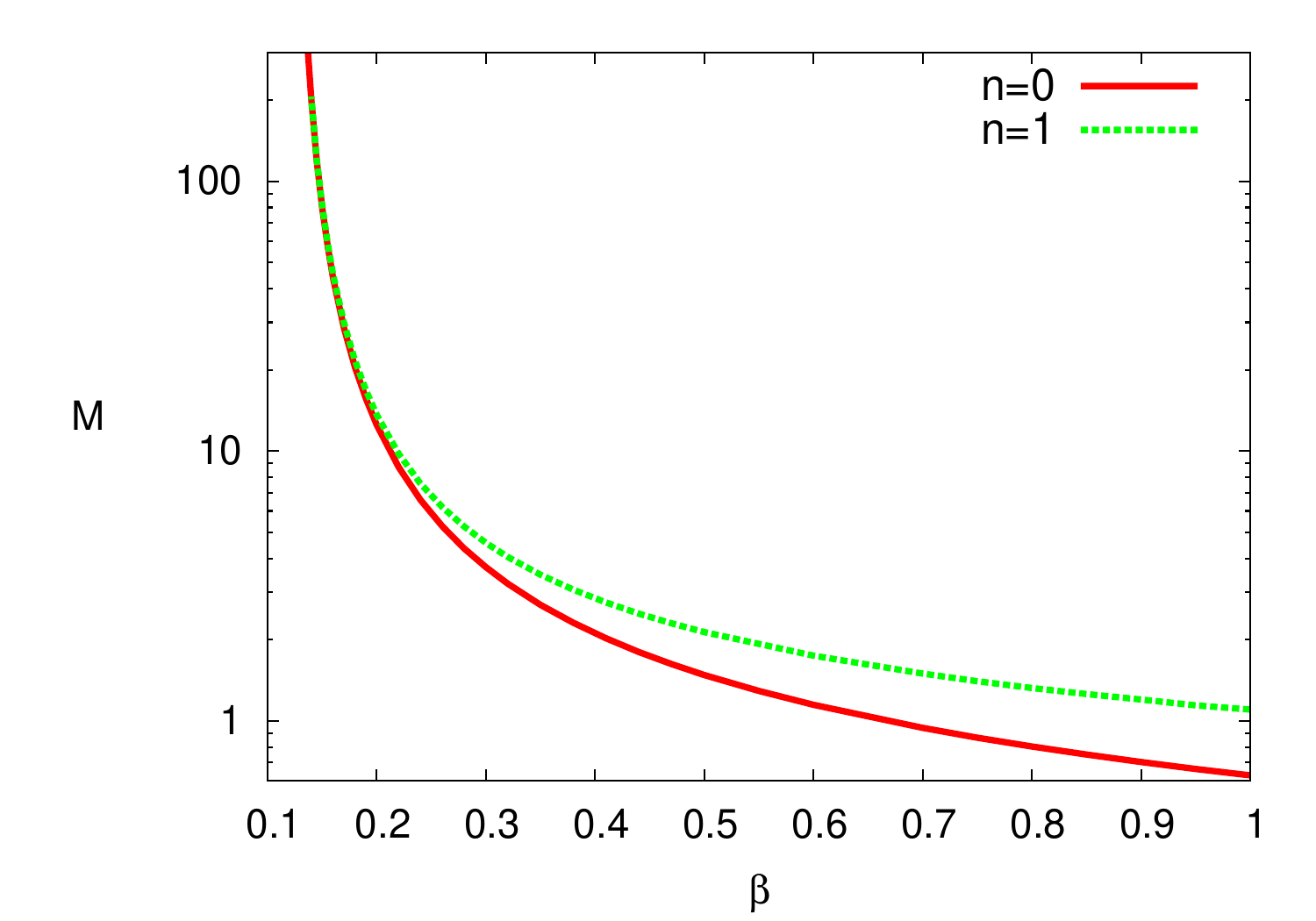}
\end{center}
\caption{\small The mass of the fundamental ($\omega=0$,
$e=0$) soliton solutions of the model
\re{Gauged_O3} with the potential \re{pot-mod}  is shown as function of the parameter $\beta$ for the spherical ($n=0$) and the axial ($n=1$)  configurations.}
    \lbfig{fig5}
\end{figure}

\begin{figure}[t!]
\begin{center}
\includegraphics[height=.31\textheight,  angle =-90]{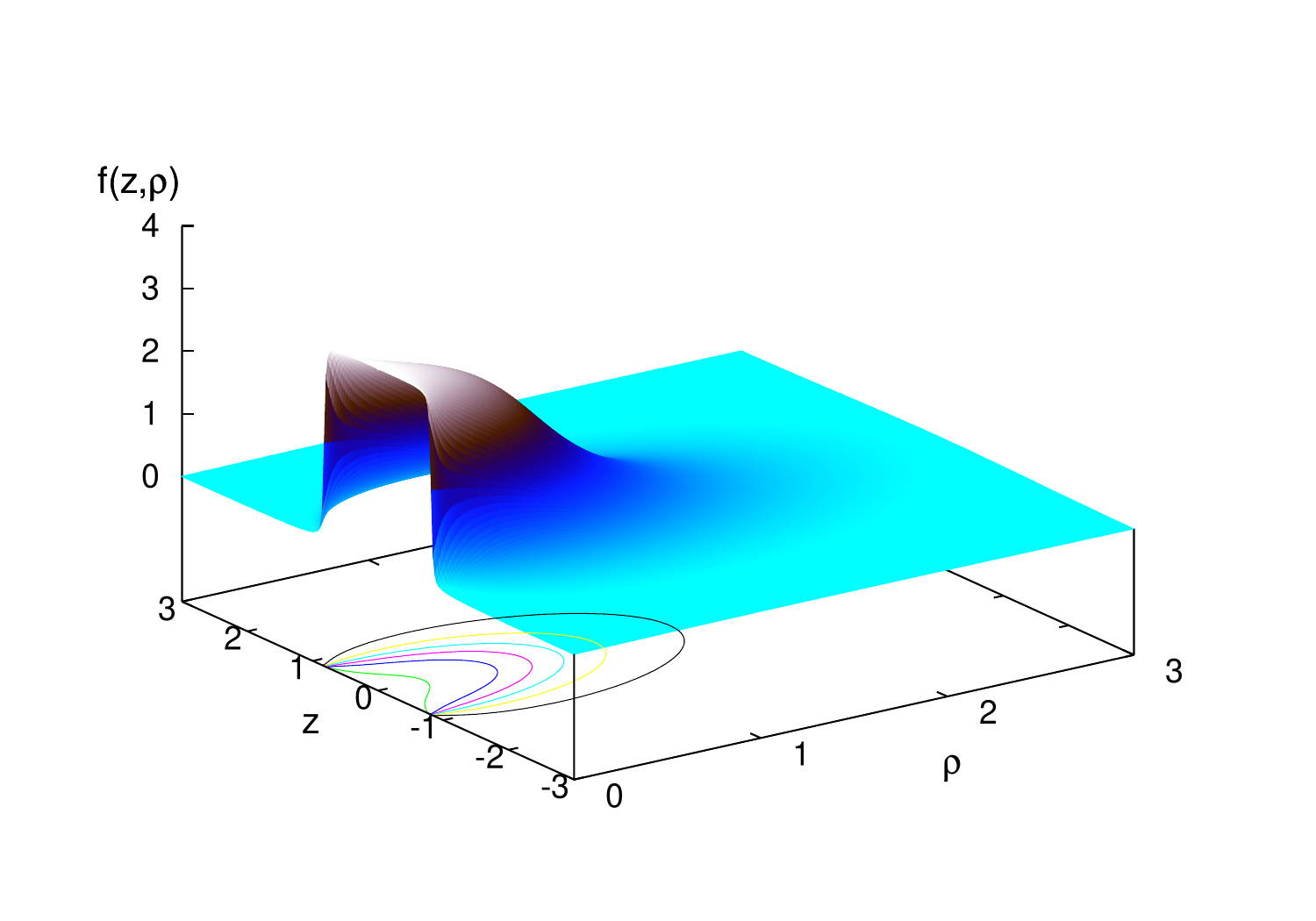}
\includegraphics[height=.32\textheight,  angle =-90]{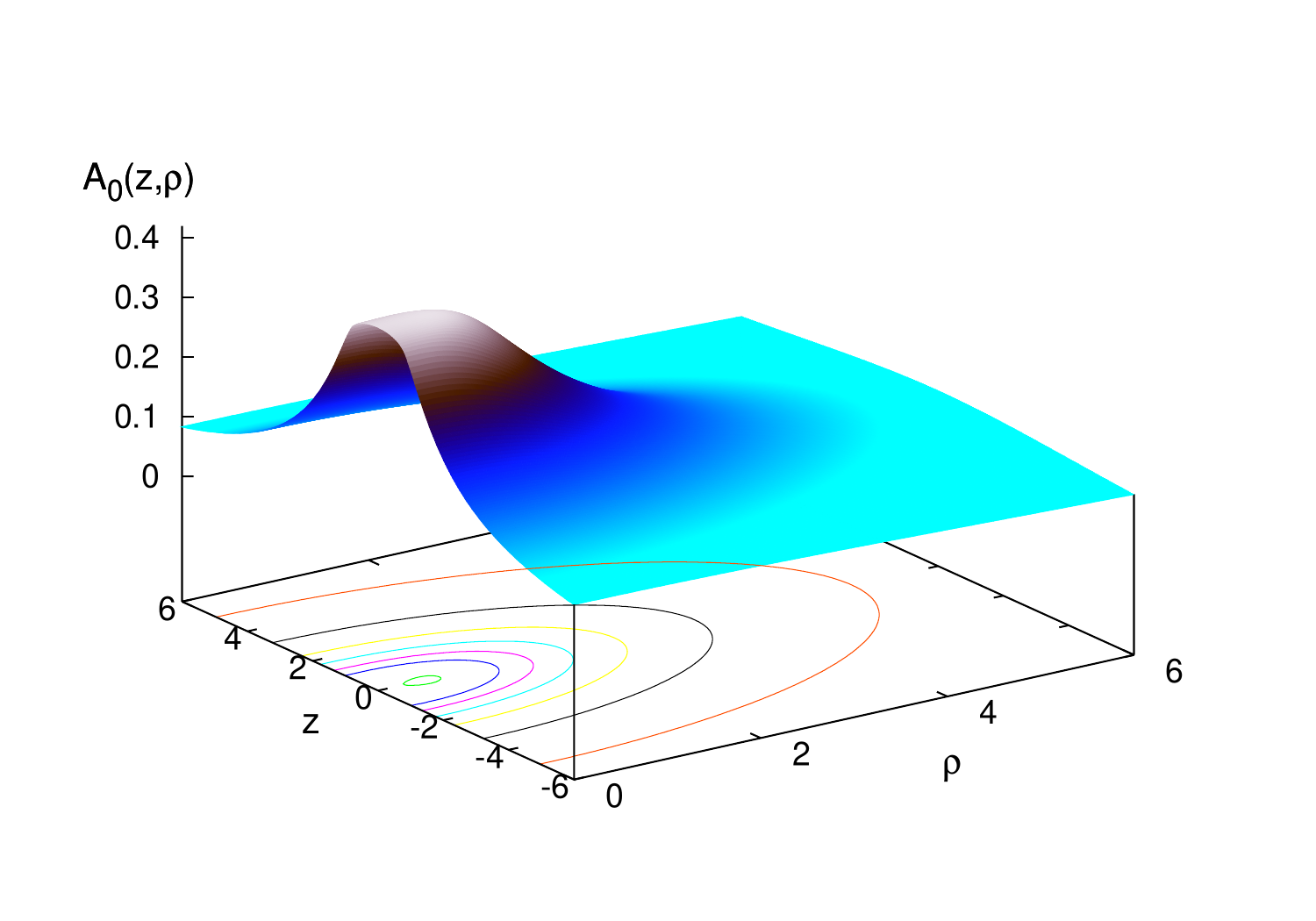}
\includegraphics[height=.32\textheight,  angle =-90]{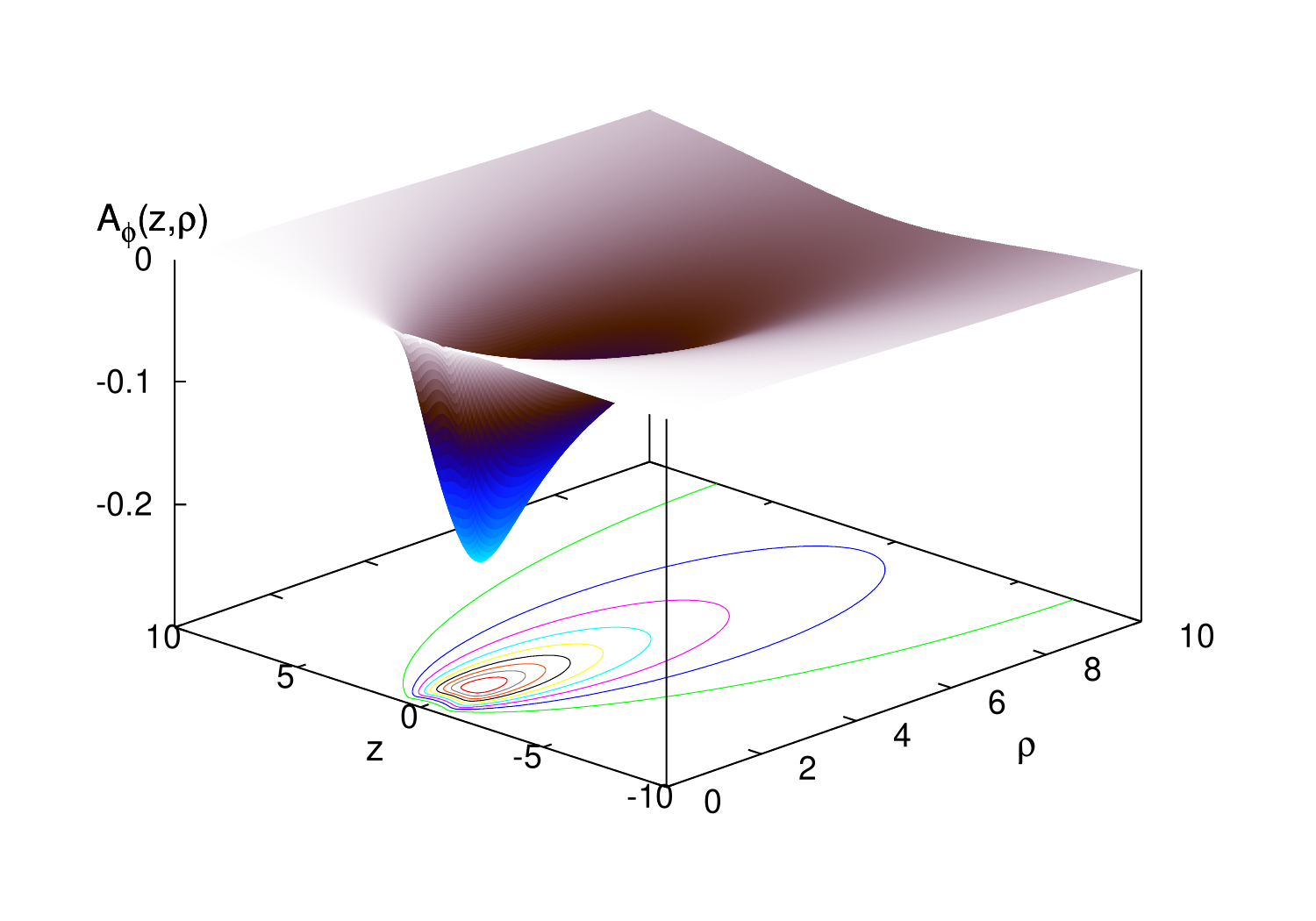}
\end{center}
\caption{\small Axially symmetric $U(1)$ gauged $O(3)$ solitons:
profiles of the scalar field amplitude  and the components of the
electromagnetic potential  are shown for an $n=1$ solution of the model
\re{Gauged_O3} with the potential \re{pot-mod} at $\omega=0.90$
and $e=0.1$. The axis are $\rho=r\sin \theta$ and $z=r\cos \theta$.
}    \lbfig{fig3}
\end{figure}

The stationary solutions are characterized by a set of physical observables, the  mass $M$,
the angular momentum $J$, the electric charge $Q_e$ and the dipole moment $\mu_m$.
Some of these  quantities can be evaluated as volume integrals
of the corresponding densities of the total  stress-energy tensor $T_{\mu\nu}=T^\phi_{\mu\nu} + T^{Em}_{\mu\nu}$,
\be
M=\int d^3 x \left( T^\mu_\mu - 2 T^t_t\right),\quad J=\int d^3 x T^t_\varphi \, , \label{MJeq}
\ee

and taking into account (\ref{Teng}), (\ref{anssfield}), (\ref{ansempot}) we can write:
\be
\begin{split}
     M&=2\pi \int_0^\pi \int_0^{\infty}r^2\sin{\theta}dr d\theta(\partial_r f^2 + \frac{\partial_\theta f^2}{r^2}+
     \frac{1}{r^2}\left(eA_\varphi+\frac{n}{\sin{\theta}}\right)^2\sin^2{f}\\
     &+\left(gA_t+\omega\right)^2\sin^2{f}+U+E_{em}),\\
     E_{em}&=\frac12\left((\partial_rA_0)^2+\frac{1}{r^2}(\partial_\theta A_t)^2+\frac{1}{r^2}(\partial_rA_{\varphi})^2+\frac{1}{r^4\sin^2{\theta}}\left(\partial_\theta(A_{\varphi}\sin{\theta})\right)^2\right).
\end{split}
\ee
\be
\begin{split}
    J&=2\pi \int_0^\pi \int_0^{\infty} r^2\sin{\theta}drd\theta\left((eA_t+\omega)(n+eA_{\varphi}\sin{\theta})\sin^2{f}+J_{em}\right),\\
    J_{em}&=\sin{\theta}\partial_rA_t\partial_rA_{\varphi}+\frac{\partial_\theta A_t(A_{\varphi}\cos{\theta}+\sin{\theta}\partial_\theta A_{\varphi})}{r^2};\label{Jmoment}
\end{split}
\ee

Also, they can be extracted from the asymptotic decay of the gauge field functions, as
\be
A_t  \rightarrow \frac{Q_e}{r} +O(\frac{1}{r^2}),\quad A_\vphi \rightarrow \frac{\mu_m \sin^2\theta}{r^2} +O(\frac{1}{r^3}) \label{asymptA}
\ee
\begin{figure}[t!]
\begin{center}
\hspace{-1.0cm}
\includegraphics[height=6.cm,angle=0,bb=00
60 365 755]{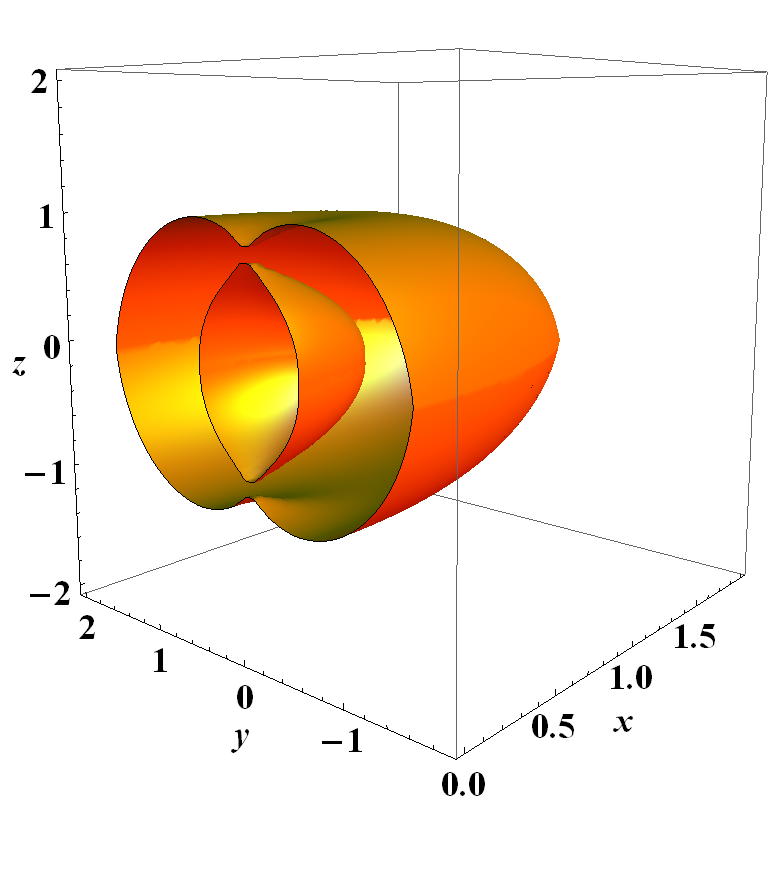}\hspace{4.0cm}
\includegraphics[height=.25\textheight,  angle =0]{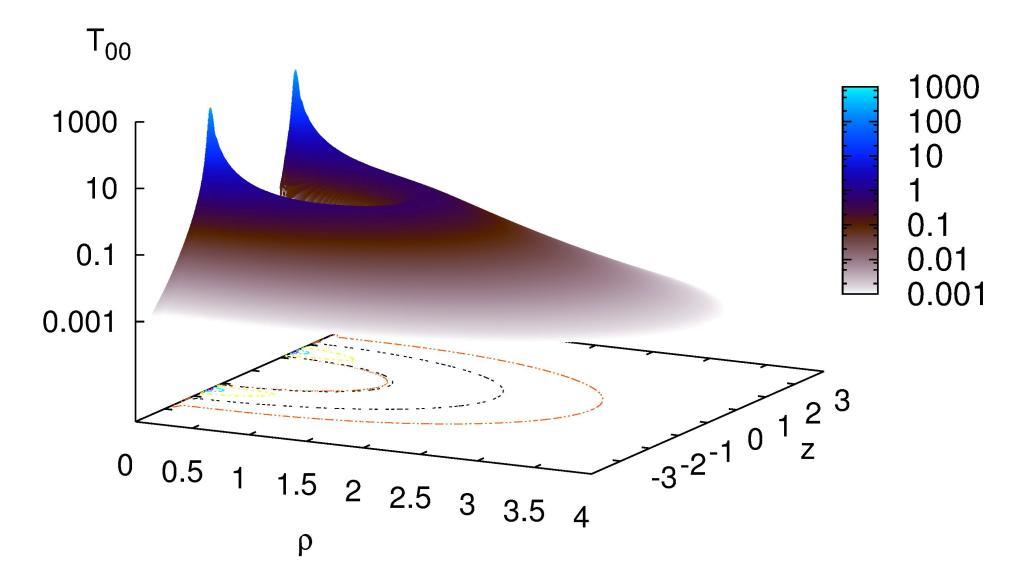}
\end{center}
\caption{\small Axially symmetric $U(1)$ gauged $O(3)$ solitons. Left plot: the
charge $Q$ isosurface at 1/5 of the maximum density of an $n=1$ solution of the model
\re{Gauged_O3} with the potential \re{pot-mod} at $ \beta =0.5, \omega=0.90$
and $e=0.1$. Right plot: Total energy density of this solution is shown as function of the coordinates $\rho=r\sin \theta$ and $z=r\cos \theta$. }
    \lbfig{fig2}
\end{figure}
The associated Noether charge of the solitons of the $O(3)$ sigma model
can be evaluated as a volume integral over
the temporal component of the conserved 4-current \re{ncurrent},
\be
    Q=\int d^3 x ~j^0=2\pi\int_0^\infty  r^2 dr\int_0^\pi\sin{\ta} d\ta\sin^2\!f
    \left(\omega+e A_t\right)\, ,
    \label{chargef}
\ee
where we make use of the parametrization of the $O(3)$ field (\ref{anssfield}).
It can be shown that $T^t_{\varphi}\sqrt{-g}=-\partial_\mu\left[\left(A_\varphi\sin{\theta}+\frac ne\right)F^{\mu4}r^2\sin{\theta}\right]$ is the total derivative, $F^{\mu4}=\left(-\partial_rA_t,-\frac{1}{r^2}\partial_\theta A_t,0,0\right)$. Also, $(T^\phi)^t_\varphi=(n+eA_\varphi\sin{\theta})j^0$.  Substituting it  into (\ref{MJeq}) and taking into account the field equations  \re{eqs-ax} together with  the boundary conditions (\ref{boundang1}), (\ref{asymptA}), one finds  that in the $U(1) $ gauged non-linear sigma model,  $J$, $Q$ and the electric charge $Q_e$ are proportional in the same way, as in the usual model with a  single complex scalar field, see e.g.
\cite{Schunck:1996he,Volkov:2002aj,Kleihaus:2005me,Collodel:2019ohy,Herdeiro:2021jgc}:
\be
\begin{split}
    J=&\int d^3 x ~T^t_\varphi=-2\pi\iint drd\theta ~\partial_\mu\left[\left(A_\varphi\sin{\theta}+\frac ne\right)F^{\mu4}r^2\sin{\theta}\right]=\\
    &-2\pi\int d\theta ~\frac ne\frac{Q_e}{r^2}r^2\sin{\theta}=-4\pi\frac{nQ_e}{e};
\end{split}
\ee
\be
\begin{split}
    Q=&2\pi\iint drd\theta~ r^2\sin{\theta}\sin^2f(\omega+eA_t)=\\
    &\frac{2\pi}{e}\iint drd\theta~ r^2\sin{\theta}\left(\partial_r^2+\frac2r\partial_r+\frac{1}{r^2}\partial_\theta^2+\frac{\cos{\theta}}{r^2\sin{\theta}}\partial_\theta\right)A_t=\\
    &-\frac{2\pi}{e}\int_0^\pi d\theta~r^2\sin{\theta}\frac{Q_e}{r^2}=-\frac{4\pi Q_e}{e}
\end{split}
\ee

Hence, \be
J = nQ =\frac{nQ_e}{e} \, ,
\ee
where $n$ is the winding number of the
scalar field. We made use of this relation to check correctness of our numerical results.

\subsection{Stationary soliton solutions}
In this section, we discuss properties of the soliton configurations. Solitons with winding number $n=1$ are constructed numerically as solutions of the set of coupled partial differential equations \re{eqs-ax}, the spherically symmetric ($n=0$) configurations are solutions of the reduced system of the radial equations \re{sys-sph-flat}. In general, the numerical calculations are performed on an
equidistant grid in spherical coordinates $r$ and $\theta$, employing
the compact radial coordinate $x=r/(C+r) \in [0:1] $ where $C$ is an arbitrary constant used to
adjust the grid according to the contraction of the solutions, and the polar angle $\theta \in [0:\pi ]$. In our numerical calculations we have
made use of a sixth-order finite difference scheme, where
the system of equations is discretized on a grid with a typical size of about 530 points in radial direction. The emerging system of nonlinear algebraic equations has been solved
using the Newton-Raphson scheme. Calculations have
been performed by employing a professional solver \cite{schoen},
with typical errors of order of $10^{-5}.$

There are a few important differences between the Q-balls in scalar theory
and  non-topological solitons of the $O(3)$ sigma model. Firstly,  usual Q-balls exist for a
finite interval of the frequency $\omega \in [\omega_{min}, \mu]$, where
a minimal allowed value of the frequency depends on the form of the potential.
In particular, the minimal  frequency can be zero,
as happens in the Fridberg-Lee-Sirlin model in Minkowski space-time \cite{Friedberg:1976me}.
In the model under consideration, there is  no lower bound on the frequency, the solutions exist
for all range of values of the angular frequency. The limiting fundamental solution at $\omega=0$
is uncharged, it corresponds to the non-topological soliton stabilized by the balance of repulsive and attractive scalar interactions \cite{Verbin:2007fa}.

\begin{figure}[t!]
\begin{center}
\includegraphics[height=.32\textheight,  angle =-90]{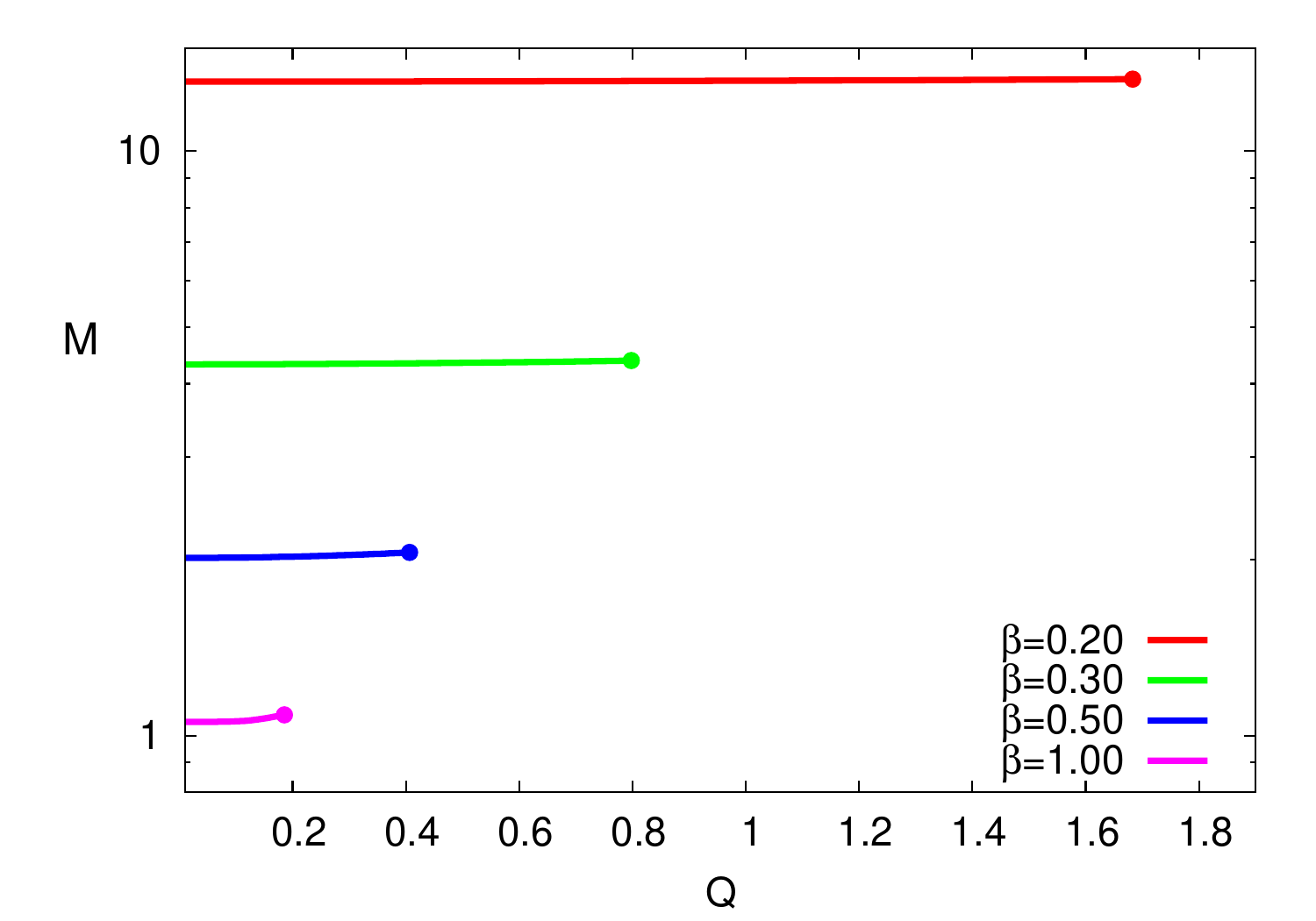}
\includegraphics[height=.32\textheight,  angle =-90]{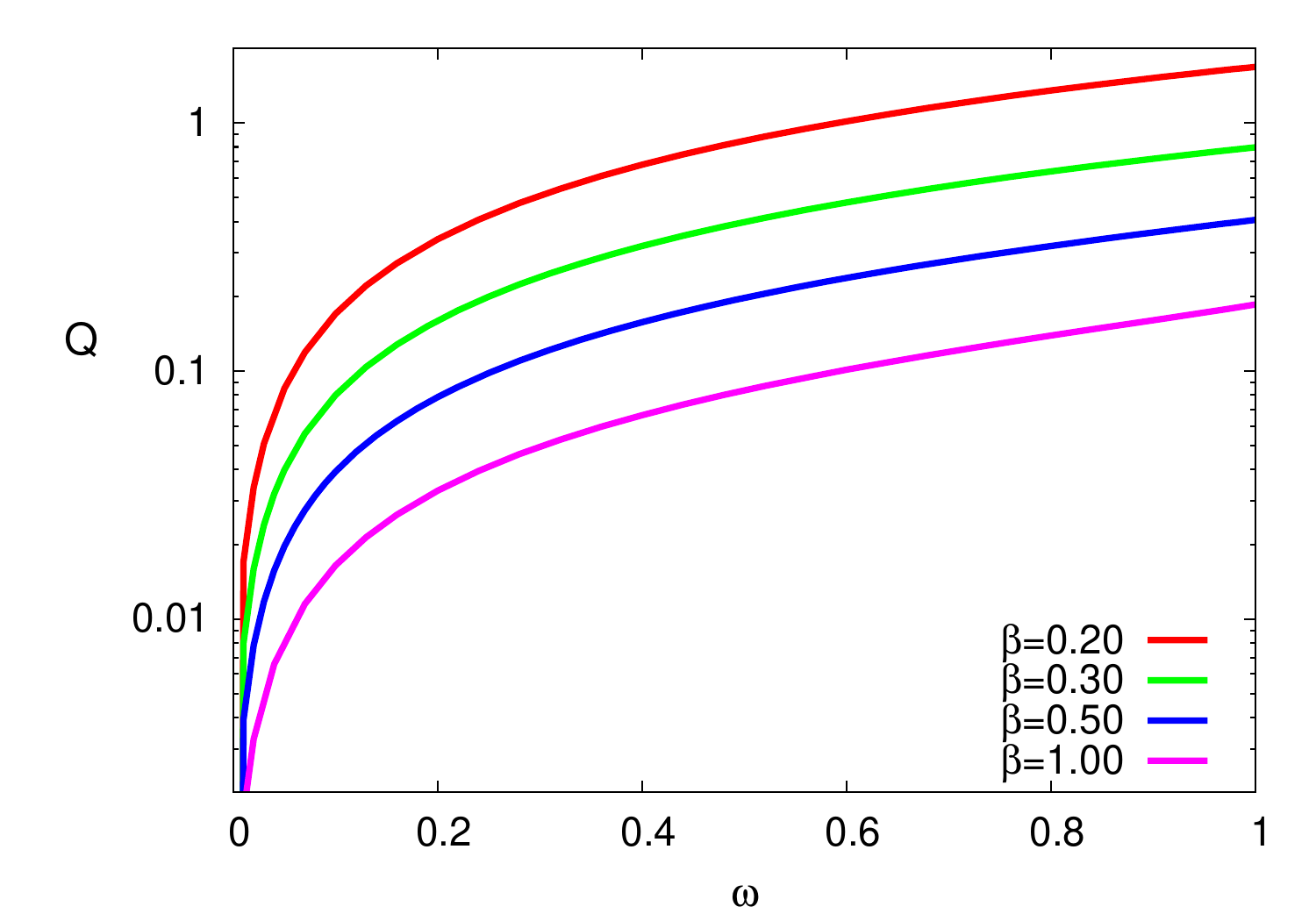}
\includegraphics[height=.32\textheight,  angle =-90]{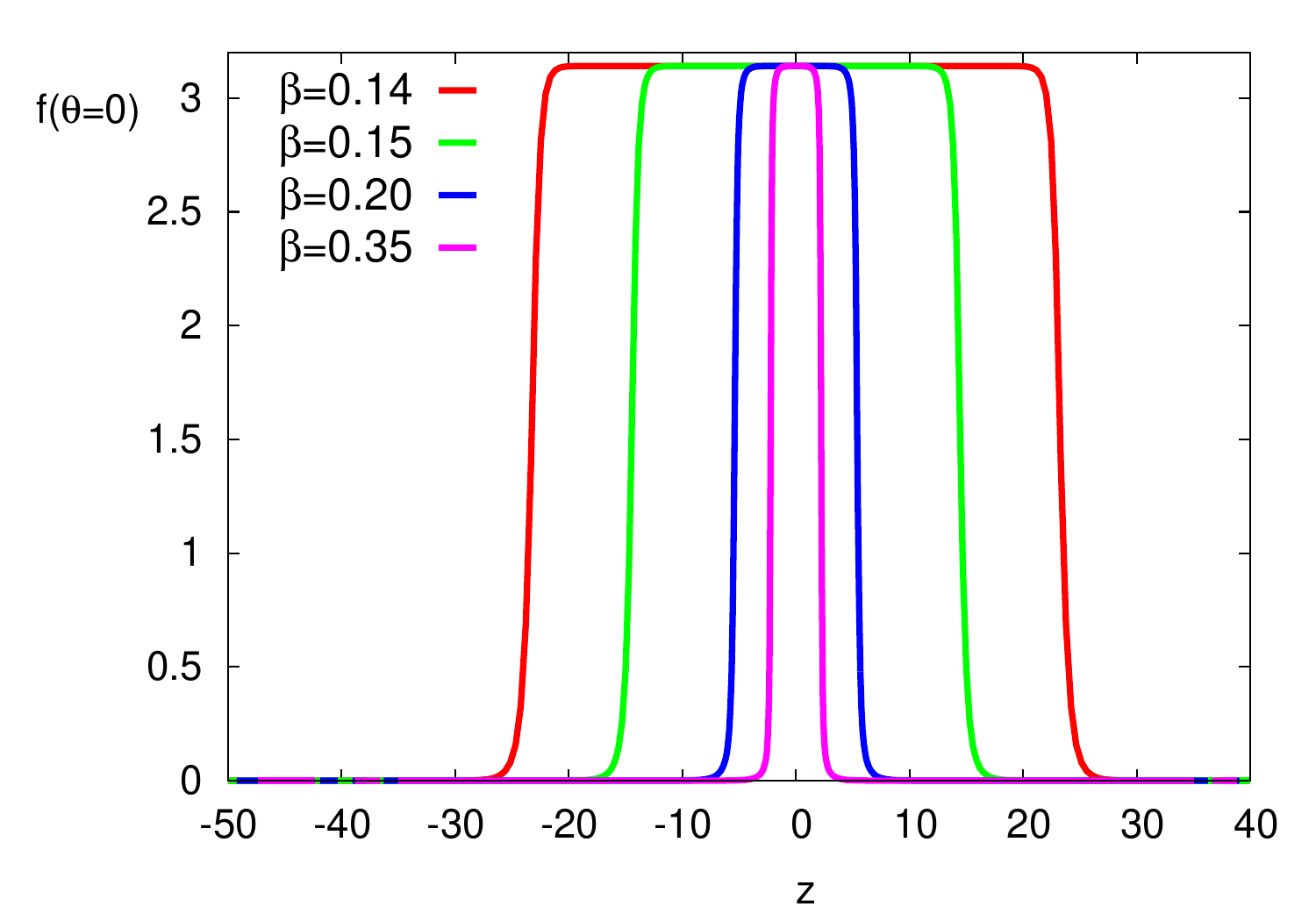}
\includegraphics[height=.32\textheight,  angle =-90]{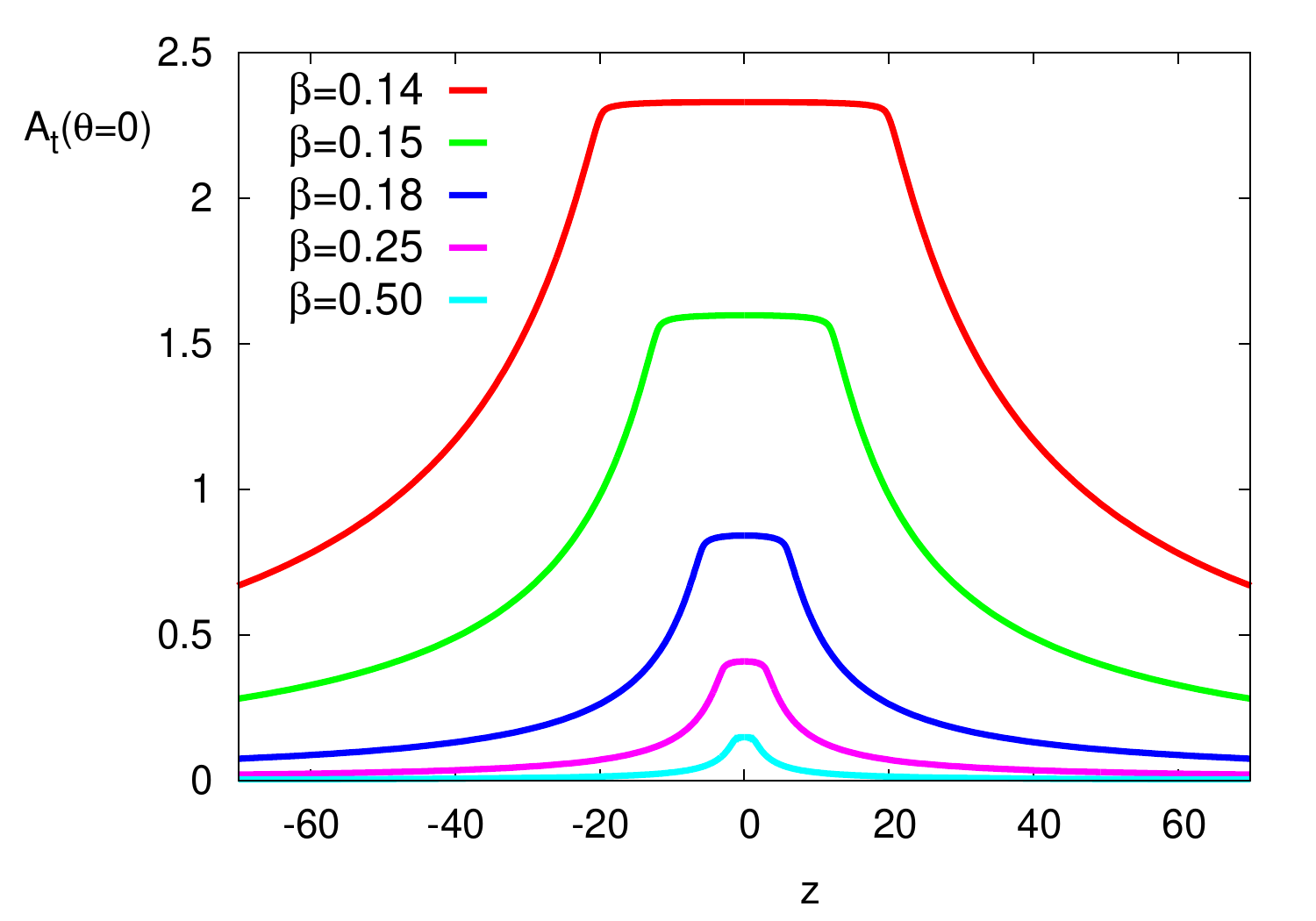}
\end{center}
\caption{\small Axially symmetric $n=1$ $U(1)$ gauged $O(3)$ solitons.
Mass vs. charge $Q$ (upper left) and the charge $Q$ vs. frequency $\omega$ (upper right)
are shown for some set of values of the parameter $\beta$ and $\mu^2=1$. The dots on the upper left plot
indicate the limiting solutions with maximal
mass and charge at the threshold $\omega=\mu$.
The profiles
of the field $f(r)$ of the fundamental ($\omega=0$,
$e=0$) solution (bottom left) and the profiles
of the field $A_t(r)$
at frequency $\omega=0.90$, and gauge coupling $e=0.1$
(bottom right)
on the symmetry axis are displayed  for some set of values of the
parameter $\beta$ of the potential \re{pot-mod} .}
    \lbfig{fig6}
\end{figure}


For both spherically ($n=0$) and axially ($n=1$) symmetric configurations there is a single branch of solutions which extends all the way up from $\omega=0$ to the mass threshold
$\mu$, the charge $Q$, as a function of the angular frequency, increases monotonically
from zero to some maximal value $Q_{max}$ as the frequency approaches the upper bound,
see Figs.~\ref{fig0},\ref{fig6} right upper plots. This value depends on the shape of the potential $\tilde U$ \re{pot-mod}.

In Fig.~\ref{fig0}, middle plots, we present the total mass $M$
of the $n=0$ solutions and the central value of the  profile function $f(0)$ versus
the angular frequency $\omega$ for a set of values of the parameter $\beta$ and fixed gauge coupling $e=0.1$. The mass of the solitons weakly increases as the frequency grows, the increase of the mass becomes more explicit for larger values of $\beta$. The central value of the function $f(0)$
decreases with increase of $\omega$, it approaches the vacuum $f(0)\sim \pi$ as  the ratio
$\beta/\mu^2$ decreases toward the critical value $1/8$.

Note that the size of the soliton also depends of the ratio of the  parameters $\beta/\mu^2$,
for a fixed value of the frequency $\omega$ it increases as the potential becomes more shallow, see Figs.~\ref{fig0},\ref{fig6} which
exhibit the profiles of the corresponding
solutions.
As
 $\beta/\mu^2 \to 1/8$
 the potential becomes infinitely degenerated,
$\tilde U(f) \xrightarrow[\beta \to 1/8]{} 2\sin^2 \frac{f}{2} (1-\sin^6\frac{f}{2})$, the interior region of all solutions, for any values of the winding number, angular frequency and the gauge coupling, turns out into a rapidly expanding ball with
the vacuum $f = \pi$ inside.
Decreasing of the parameter $\beta$ increases the
mass of the configurations, as seen in the left upper plot of Figs.~\ref{fig0},\ref{fig6}. The mass of the solutions diverge in the limit
$\beta/\mu^2 \to 1/8$, see Fig.~\ref{fig5}.

Secondly, the Noether charge  $Q$ \re{chargef} cannot be interpreted as the particle number, the
norm of the $O(3)$ field is fixed and only two components
of the triplet contribute to the charge \re{chargef}. In Fig.~\ref{fig2} we displayed an isosurface of the charge
density distribution of an illustrative $n=1$ solution. Unlike the corresponding distribution of the charge density of an
axially-symmetric Q-ball in a scalar theory, which represent a torus, the charge distribution of the $O(3)$ soliton has
a shell-like structure, as seen in Fig.~\ref{fig2}, left plot. This is related with the form of the profile functions
which are shown in Fig.~\ref{fig3}. The central value of the profile function interpolates between $f\left.\right|_0 \sim \pi$ and $f\left.\right|_\infty =0$, thus the  charge density \re{chargef} has a minimum at the center  of the soliton and it is vanishing on the spacial infinity with a maximum at the wall separating the vacua.

Indeed, the amplitude of the scalar field feature a plateau around the center at which a false vacuum
is formed, see left bottom plot of Figs.~\ref{fig0},\ref{fig6}. The plateau ends with two steps on the symmetry axis, where the field $f$ drops to the true vacuum. The energy density
distribution has two sharp peaks at these points, as seen in Fig.~\ref{fig2}, right plot. Notice that the electric potential $A_t$ is a constant in the interior of the soliton, as displayed in the Figs.~\ref{fig0}, \ref{fig6} bottom right plots. Thus, the electric field of the gauged configuration is vanishing inside. Outside the core of the soliton it possesses a
Coulomb-like asymptotic tail . It corresponds to the long-range field of the charge $Q_e=eQ$.  On the other hand, the magnetic field of the $n=1$ gauged soliton is toroidal, as in the case of
the usual gauged Q-balls.

Thirdly, in the case of the usual ungauged Q-balls in a complex scalar theory, the size of the soliton
grows indefinitely as the angular frequency approaches the lower minimal value,
the value of the field amplitude $f$ at the origin rapidly increases and
both the mass and the charge of the configuration diverge in this limit.

In the case of solitonic solutions of the $O(3)$ model, the maximum of the
field amplitude $f(0)$  is restricted by the constrain. When $\omega$ is
decreased, it slowly approaches its maximal value which remains slightly below $f_{cr}=\pi$,
even in limit $\omega \to 0$, while
the interior region of the soliton is slowly increasing. Thus, in this limit the
mass of the configuration remains finite albeit the Noether charge $Q\sim \omega$ is vanishing.

Fourthly, the usual evolution pattern of the gauged Q-balls in a complex scalar theory
is that there are two branches of solutions, first of which arises from the
pertubative excitations as the angular frequency is decreasing below the mass threshold
\cite{Lee:1988ag,Anagnostopoulos:2001dh,Gulamov:2013cra,Gulamov:2015fya}. A
bifurcation with the second, upper in energy branch, occurs at some minimal non-zero value
of the angular frequency. The second branch of gauged
Q-balls extends forward up to the maximal value of the frequency, along this  branch the
characteristic size of the gauged Q-balls rapidly
increases, it inflates as $\omega$ approaches its
upper critical value. The minimal critical value of the frequency increases with the increase of the
gauge coupling $e$, an additional repulsive interaction arises from the gauge
sector and the solutions cease to exist at some maximal critical value of the gauge coupling.

On the contrary, there is only one branch of charged Q-balls in the $O(3)$ sigma model,
increase of the gauge coupling affects the value of the Noether charge and drives the configurations
closer to the limit $f_{cr}=\pi$. However, there is no second branch of solutions for any values of the gauge
coupling and the parameters of the potential $\mu$ and $\beta$.

Fifthly,  in the theory of a complex scalar field, Q-balls typically arise smoothly from perturbative
excitations about the vacuum, as the angular frequency is decreasing below the mass threshold,
and scalar quanta condense into a non-topological soliton.  It is not the case for the solitons of the $O(3)$ sigma model,
they are disconnected from the vacuum excitations for most of the range of values of the potential parameter $\beta$.

Finally, we note that stable soliton solutions of the model do not exist as the ratio $\beta/\mu^2$ becomes smaller
than $1/8$, so there is no flat space solutions
in the gauged $O(3)$ sigma model with the pion mass potential.

\section{Conclusions}

The main purpose of this paper is to show the existence of new type of
non-topological soliton solutions of the $U(1)$ gauged non-linear $O(3)$ sigma model in 3+1 dimensional flat space-time. Unlike usual Q-balls
in a complex field theory, they are stabilized by the special choice of the symmetry breaking potential rather than isorotations.
Considering dependency of the solutions on the frequency, we observe only one branch of charged Q-balls in the $O(3)$ sigma model, is originates from static ($\omega=0$) configuration and extends up to the mass threshold.

The solitons of the $U(1)$ gauged  non-linear $O(3)$ sigma model exhibit examples of the configurations with a quantized angular momentum, $J=nQ$ and with both the
electric charge and toroidal magnetic field, which forms a vortex encircling
the soliton.

The work here should be taken further by considering a family of potentials which would support non-zero vacuum expectation value of the scalar field. It may allow to introduce Higgs-like mechanism providing a mass to the gauge field. Another direction can be related with investigation of properties of the self-gravitating soltions in the $U(1)$ gauged  non-linear $O(3)$ sigma model minimally coupled to Einstein gravity. We hope to address these problems in our future work.

\section*{Acknowledgements}
Y.S. thanks Eugen Radu
for useful discussions. He also gratefully acknowledges the support  by FAPESP,
project No 2024/01704-6 and thanks the Instituto de F\'{i}sica de S\~{a}o Carlos; IFSC for kind hospitality. L.A.F. is partially support by the CNPq grant 307833/2022-4.


\begin{small}

\end{small}

\begin{thebibliography}{99}
\bibitem{Gell-Mann:1960mvl}
M.~Gell-Mann and M.~Levy,
Nuovo Cim. \textbf{16} (1960), 705
\bibitem{Manton:2004tk}
  N.~S.~Manton and P.~Sutcliffe,
  {\it 'Topological solitons',}
    Cambridge University Press, 2004.
\bibitem{Shnir:2018yzp}
Y.~M.~Shnir,
{\it 'Topological and Non-Topological Solitons in Scalar Field Theories,'}
Cambridge University Press, 2018
\bibitem{Polyakov:1975yp}
A.~M.~Polyakov and A.~A.~Belavin,
JETP Lett. \textbf{22} (1975), 245-248
\bibitem{Derrick:1964ww}
G.~H.~Derrick,
J. Math. Phys. \textbf{5} (1964), 1252-1254
\bibitem{Nicole:1978nu}
D.~A.~Nicole,
J. Phys. G \textbf{4} (1978), 1363
\bibitem{Faddeev:1976pg}
L.~D.~Faddeev,
Lett. Math. Phys. \textbf{1} (1976), 289
\bibitem{Foster:2010zb}
D.~Foster,
Phys. Rev. D \textbf{83} (2011), 085026
\bibitem{Shnir:2014mfa}
Y.~Shnir and G.~Zhilin,
Phys. Rev. D \textbf{89} (2014) no.10, 105010
\bibitem{Samoilenka:2018oil}
A.~Samoilenka and Y.~Shnir,
Phys. Rev. D \textbf{97} (2018) no.12, 125014
\bibitem{Leese:1991hr}
R.~A.~Leese,
Nucl. Phys. B \textbf{366} (1991), 283
\bibitem{Ward:2003un}
R.~S.~Ward,
J. Math. Phys. \textbf{44} (2003), 3555
\bibitem{Harland:2013uk}
D.~Harland, J.~J\"aykk\"a, Y.~Shnir and M.~Speight,
J. Phys. A \textbf{46} (2013), 225402
\bibitem{Battye:2013xf}
R.~A.~Battye and M.~Haberichter,
Phys. Rev. D \textbf{87} (2013) no.10, 105003
\bibitem{Friedberg:1976me} R.~Friedberg, T.D.~Lee and A.~Sirlin,
Phys.\ Rev.\ D {\bf 13} (1976)  2739.
\bibitem{Rosen:1968mfz}
G.~Rosen,
J. Math. Phys. \textbf{9} (1968), 996
\bibitem{Coleman:1985ki}S.R.~Coleman,
Nucl.\ Phys.\ B  {\bf 262}  (1985)  263; Erratum: Nucl.\ Phys.\ B  {\bf 269} (1986)   744
\bibitem{Verbin:2007fa}
Y.~Verbin,
Phys. Rev. D \textbf{76} (2007), 085018
\bibitem{Herdeiro:2018djx}
C.~Herdeiro, I.~Perapechka, E.~Radu and Y.~Shnir,
JHEP \textbf{02} (2019), 111
\bibitem{Cano:2023bpe}
P.~A.~Cano, L.~Machet and C.~Myin,
Phys. Rev. D \textbf{109} (2024) no.4, 044043
\bibitem{Adam:2025ktm}
C.~Adam, J.~C.~Mourelle, A.~Garc\'\i{}a Mart\'\i{}n-Caro and A.~Wereszczynski,
[arXiv:2502.20923 [gr-qc]].
\bibitem{Kleihaus:2007vk}
  B.~Kleihaus, J.~Kunz, M.~List and I.~Schaffer,
  Phys.\ Rev.\ D {\bf 77} (2008) 064025
\bibitem{Lee:1988ag}
K.~M.~Lee, J.~A.~Stein-Schabes, R.~Watkins and L.~M.~Widrow,
Phys. Rev. D \textbf{39} (1989), 1665
\bibitem{Anagnostopoulos:2001dh}
K.~N.~Anagnostopoulos, M.~Axenides, E.~G.~Floratos and N.~Tetradis,
Phys. Rev. D \textbf{64} (2001), 125006
\bibitem{Gulamov:2013cra}
I.~E.~Gulamov, E.~Y.~Nugaev and M.~N.~Smolyakov,
Phys. Rev. D \textbf{89} (2014) no.8, 085006
\bibitem{Gulamov:2015fya}
I.~E.~Gulamov, E.~Y.~Nugaev, A.~G.~Panin and M.~N.~Smolyakov,
Phys. Rev. D \textbf{92} (2015) no.4, 045011
\bibitem{Schroers:1995he}
B.~J.~Schroers,
Phys. Lett. B \textbf{356} (1995), 291-296
\bibitem{Amari:2024adu}
Y.~Amari, M.~Eto and M.~Nitta,
JHEP \textbf{11} (2024), 127
\bibitem{Loginov:2016yeh}
A.~Y.~Loginov,
Phys. Rev. D \textbf{93} (2016) no.6, 065009
\bibitem{Ghosh:1995ze}
P.~K.~Ghosh and S.~K.~Ghosh,
Phys. Lett. B \textbf{366} (1996), 199-204
\bibitem{Samoilenka:2015bsf}
A.~Samoilenka and Y.~Shnir,
Phys. Rev. D \textbf{93} (2016) no.6, 065018
\bibitem{Leese:1989gi}
R.~A.~Leese, M.~Peyrard and W.~J.~Zakrzewski,
Nonlinearity \textbf{3} (1990), 773-808
\bibitem{Salmi:2014hsa}
P.~Salmi and P.~Sutcliffe,
J. Phys. A \textbf{48} (2015) no.3, 035401
\bibitem{Gillard:2015eia}
M.~Gillard, D.~Harland and M.~Speight,
Nucl. Phys. B \textbf{895} (2015), 272-287
\bibitem{Lee:1991ax}
T.~D.~Lee and Y.~Pang,
Phys. Rept. \textbf{221} (1992), 251-350
\bibitem{Battye:2014qva}
R.~A.~Battye, M.~Haberichter and S.~Krusch,
Phys. Rev. D \textbf{90} (2014) no.12, 125035
\bibitem{Ioannidou:2006nn}
T.~Ioannidou, B.~Kleihaus and J.~Kunz,
Phys. Lett. B \textbf{643} (2006), 213-220
\bibitem{Perapechka:2017bsb}
I.~Perapechka and Y.~Shnir,
Phys. Rev. D \textbf{96} (2017) no.12, 125006
\bibitem{Herdeiro:2018daq}
C.~Herdeiro, I.~Perapechka, E.~Radu and Y.~Shnir,
JHEP \textbf{10} (2018), 119
\bibitem{Schunck:1996he}
F.~E.~Schunck and E.~W.~Mielke,
Phys. Lett. A \textbf{249} (1998), 389-394
\bibitem{Volkov:2002aj}M.S.~Volkov and E.~Wohnert,
Phys.\ Rev.\  D  {\bf 66} (2002)  085003.
\bibitem{Kleihaus:2005me}B.~Kleihaus, J.~Kunz and M.~List,
Phys.\ Rev.\ D  {\bf 72} (2005)  064002 .
\bibitem{Collodel:2019ohy}
L.G.~Collodel, B.~Kleihaus and J.~Kunz,
Phys. Rev. D \textbf{99} (2019) no.10, 104076
\bibitem{Herdeiro:2021jgc}
C.~Herdeiro, I.~Perapechka, E.~Radu and Y.~Shnir,
Phys. Lett. B \textbf{824} (2022), 136811
\bibitem{schoen}
W.~Sch\"onauer  and R.~Wei\ss,
``Efficient vectorizable PDE solvers"
J. Comput. Appl. Math. 1989. V. 27. P. 279;
M.~Schauder, R.~Wei\ss , and  W.~Sch\"onauer,
``The CADSOL Program Package",
Universit\"at Karlsruhe, 1992. Interner Bericht Nr. 46/92






\end{thebibliography}
\end{document}